\documentclass[12pt]{article}
\usepackage[dvips]{color}
\usepackage{epsfig}
\usepackage{amsmath}
\usepackage{graphicx}

\begin{document}
\title{\bf Hubble parameter corrected interactions in cosmology}
\author{{J. Sadeghi$^{a}$ \thanks{Email: pouriya@ipm.ir},\hspace{1mm} M. Khurshudyan$^{b}$ \thanks{Email:
khurshudyan@yandex.ru},\hspace{1mm} M. Hakobyan$^{c,d}$ \thanks{Email: margarit@mail.yerphi.am}\hspace{1mm}
and H. Farahani$^{a}$ \thanks{Email:
h.farahani@umz.ac.ir}}\\
$^{a}${\small {\em Department of Physics, Mazandaran University, Babolsar, Iran}}\\
{\small {\em P.O.Box 47416-95447, Babolsar, Iran}}\\
$^{b}${\small {\em Department of Theoretical Physics, Yerevan State
University,}}\\
{\small {\em 1 Alex Manookian, 0025, Yerevan, Armenia}}\\
$^{c}${\small {\em A.I. Alikhanyan National Science Laboratory,
Alikhanian Brothers St., Yerevan, Armenia} } \\
$^{d}${\small {\em Department of Nuclear Physics, Yerevan State University, Yerevan, Armenia}}}  \maketitle
\begin{abstract}
In this paper we make steps in a new direction by considering fluids with EoS of more general form $F(\rho,P)=0$. It is thought that there should be interaction between cosmic fluids, but this assumption for this stage carries only phenomenological character opening a room for different kind of manipulations. In this article we will consider a modification of an interaction $Q$, where we accept that interaction parameter $b_{1}$ (order of unity) in $Q=3Hb_{1}\rho$ is time dependent and presented as a linear function of Hubble parameter $H$ of the form $b_{0}+btH$, where $b$ and $b_{0}$ are constants. We consider two different models include modified Chaplygin gas and palotropic gas which have bulk viscosity. Then, we investigate problem numerically and analyze behavior of different cosmological parameters concerning to fluids and behavior of Universe.
\end{abstract}

\section{\large{Introduction}}
Experimental data interpretation claims that we have accelerated expansion for our Universe. However this phenomenon can be understood as a theoretical model based consequence. In general relativity concepts of dark energy and dark matter were introduce by hand and seems that they deal with the problem at intermediate level, because considered number of models and articles is going to be behind of reasonable limit. However still questions concerning to the nature of dark energy, dark mater, about possible interactions etc are open. Dark energy thought to be responsible to accelerated expansion. On theoretical and phenomenological level scalar fields were considered as thought that scalar field can be a base of dark energy. One of them is a Tachyonic scalar field. Concerning to some fundamental problems dynamical models of dark energy were proposed and considered from different corners. However, it is not the unique approach and geometrical part of gravitational action were modified.\\
A set of observational data reveal the following picture of our
Universe called modern era in theoretical cosmology, that an expansion of our Universe is accelerated [1-3]. Then, the density of matter is very much less than critical density [4], the Universe is flat and the total energy density is very close to the critical [5]. Explanation of accelerated expansion of our Universe takes two different ways and now they are developing and evaluating as different approaches,
however there is not any natural restriction of a possibilities of recombination of two approaches in one single approach. In that case we believe that joined approach will be more sufficient and rich with new and interesting physics. To explain recent observational data, which reveals accelerated expansion character of the Universe, several models were proposed. One of the possible scenarios (general relativity framework) is the existence of a dark energy ($73\%$ of the energy of our Universe) with negative pressure and positive energy density giving an acceleration to the expansion [6, 7]. Other component, dark matter occupies about $23\%$, and usual baryonic matter occupies about $4\%$. Among different viewpoints concerning to the nature of the dark component of the Universe, we would like to mention a scalar field models, one of them is Tachyonic field with its relativistic Lagrangian,
\begin{equation}\label{eq:tach lag}
L_{TF}=-V(\phi)\sqrt{1-\partial_{\mu}\phi\partial^{\nu}\phi},
\end{equation}
which captured a lot of attention (see, for instance, references in \cite{Murli}).
The stress energy tensor,
\begin{equation}\label{eq:energy tensor}
T^{ij}=\frac{\partial L}{\partial (\partial_{i}\phi)}\partial^{j}\phi-g^{ij}L,
\end{equation}
gives the energy density and pressure as,
\begin{equation}\label{eq:tachyonic density}
\rho=\frac{V(\phi)}{\sqrt{1- \partial_{i}\phi \partial^{i}\phi}},
\end{equation}
and,
\begin{equation}\label{eq:tachyonic pressure}
P=-V(\phi)\sqrt{1- \partial_{i}\phi \partial^{i}\phi}.
\end{equation}
A quintessence field [16] is other model based on scalar field with standard kinetic term, which minimally coupled to gravity. In that case the action has a wrong sign kinetic term and the scalar field is called phantom [17]. Combination of the quintessence and the phantom is known as the quintom model [18]. Extension of kinetic term in Lagrangian yields to a more general frame work on field theoretic
dark energy, which is called k-essense [19, 20]. A singular limit of k-essense is called Cuscuton model [21].
This model has an infinite propagating speed for linear perturbations, however causality is still valid. The
most general form for a scalar field with second order equation of motion is the Galileon field which also
could behaves as dark energy [22].\\
Dark energy models based on idea of fluid do not less popular and well studied. Fluids in cosmology are convenient, because as practice teach us, we can for instance, different modifications in geometrical part of action encode in fluid part of field equations, giving illusion that in nature fluids with general form of EoS could be considered like to Chaplygin gas and its generalizations [9-15]. There are several models to describe dark energy such as the cosmological constant and its generalizations [8]. Among various models of dark energy, a new model of dark energy called Veneziano ghost dark (GD) energy, which supposed to exist to solve the $U(1)_{A}$ problem in low-energy effective theory of QCD, and has attracted a lot of interests in recent years \cite{Urban0}-\cite{Chao2}. Indeed, the contribution of the ghosts field to the vacuum energy in curved space or time-dependent background can be regarded as a possible candidate for the dark energy.
It is completely decoupled from the physics sector. Veneziano ghost is unphysical in the QFT formulation
in Minkowski space-time, but exhibits important non trivial physical effects in the expanding Universe. It is hard to accept
such linear behavior and it is thought that there should be some exponentially small corrections. However,
it can be argued that the form of this behavior can be result of the fact of the very complicated topological
structure of strongly coupled QCD. This model has advantage compared to other models of dark energy,
which can be explained by standard model and general relativity. Comparison with experimental data, reveal
that the current data does not favorite compared to the $\Lambda$CDM model, which is not conclusive and future
study of the problem is needed. Energy density of ghost dark energy may reads as,
\begin{equation}\label{eq:GDE}
\rho_{GD} = \theta H,
\end{equation}
where $H$ is Hubble parameter  and $\theta$ is constant parameter of the model, which should be determined.
The relation (5) generalized by the Ref. \cite{Cai} as the following,
\begin{equation}\label{eq:GDE}
\rho_{GD} = \theta H +\vartheta H^{2},
\end{equation}
where $\theta$ and $\vartheta$ are constant parameters of the model. Such kind of fluids could be named as a geometrical
fluids, because it is clear that it contains information about geometry of the space-time and metric. Recently
a model of varying ghost dark energy were proposed in the Ref. \cite{Khurshudyan3} and extended to the case of interaction with variable $\Lambda$ and $G$ [28]. Unfortunately, the pure models based on the energy density (5) and (6) may ruled out.
This has been recently shown in detail in [39] from the point of view of cosmic perturbations, but it was already indicated in the previous works [40, 41]. It bring us to consider some corrections such as viscosity and interaction to obtain valid model.
Moreover, there is another problem with dark energy models of the form (5) and (6) which is they do not have a $\Lambda$CDM limit. The absence of an additive term in the structure of the dark energy is highly problematic as the aforementioned works show. Irrespective of the theoretical motivations for these models the bare truth is that they are phenomenologically excluded. But there is also a fundamental motivation raising theoretical doubts on these models. The existence of linear terms in the Hubble rate is incompatible with the general covariance of the effective action of QFT in curved space-time. This is mentioned also in the Refs. [39-41] but it is discussed in more detail in Refs. [42, 43], where it is also shown how to correctly generalize these models for the physics of the early universe by including only even powers of the Hubble rate. However, another corrections like including interaction term and viscosity may resolve above problems.\\
In this article we would like to propose a modification in interaction $Q=3Hb_{1}\rho$ which by the general idea should be exist between cosmic components. We will assume the interaction term as the following,
\begin{equation}\label{eq:interQ}
Q=3H(b_{0}+btH)\rho,
\end{equation}
where $b_{0}$ and $b$ are constants. This assumption will bring us to the possibility that $b_{1}$ is a function of time. Such assumption already were considered in Ref. \cite{Chen}, while in Ref. \cite{Khurshudyan3} interacting varying Ghost DE models were considered with time dependent interaction term. Assumption were, that $b(t)=a(t)^{\xi}$.\\
Due to the lack of information about dark energy and dark matter, usually the interaction terms assumed to be proportional to
the energy density, scale factor, Hubble parameter and their derivative. In the Ref. [44] the general time-dependent interactions considered which proves
that even very simple forms can alleviate the coincidence problem, and lend the cosmic
acceleration a transient character. This makes a good motivation to consider time-dependent interaction of the form the equation (7).\\
Before to main formulation of our problem we would like to pay our attention to the question of interaction in cosmology between fluid components. Usually, three forms of $Q$ are used,
\begin{equation}\label{eq:Q1}
Q=3Hb_{1}\rho_{de},
\end{equation}
\begin{equation}\label{eq:Q2}
Q=3Hb_{1}(\rho_{de}+\rho_{dm}),
\end{equation}
and,
\begin{equation}\label{eq:Q2}
Q=3Hb_{1}\rho_{dm},
\end{equation}
where $b_{1}$ is a coupling constant. From the thermodynamical view, it is argued that the second law of thermodynamics strongly favors that dark energy decays into dark matter, which implies $b$ to be positive. These type of interactions are either positive or negative and can not change sign. However, recently by using a model independent method to deal with the observational data Cai and Su found that the sign of interaction $Q$ in the dark sector changed in the redshift range of $0.45 \leq z \leq 0.9$.  Hereafter, a sign-changeable interaction \cite{Hao}-\cite{SChint1} were introduced,
\begin{equation}\label{eq:signcinteraction}
Q=q(\alpha\dot{\rho}+3\beta H\rho).
\end{equation}
where $\alpha$ and $\beta$ are dimensionless constants, the energy density $\rho$ could be $\rho_{dm}$, $\rho_{\small{de}}$, $\rho_{tot}$. $q$ is the deceleration parameter given by,
\begin{equation}\label{eq:decparameter}
q=-\frac{1}{H^{2}} \frac{\ddot{a}}{a}=-1-\frac{\dot{H}}{H^{2}}.
\end{equation}
This new type of interaction, where deceleration parameter $q$ is a key ingredient makes this type of interactions different from the ones considered in literature and presented above, because it can change its sign when our Universe changes from deceleration $q>0$ to acceleration $q<0$. $\gamma \dot{\rho}$ is introduced from the dimensional point of view. We would like also to stress a fact, that by this way we import a more information about the geometry of the Universe into the interaction term. This fact can be mean that we should consider more general forms for the interaction term. It is obvious that this splitting (as a mathematical act) can be done for any fluid with any number of components making a linear combination of pressure and energy density. From equations we see that unit of interaction $Q$ should be $time^{-1}\times energy~density$. Other type of interaction is of the form $Q=\gamma\dot{\rho}$, where for $\rho$ we can say the same as in previous case. Question of $time^{-1}$ here was solved by taking derivative of energy density instead of using Hubble parameter with $time^{-1}$ unit. Combination of these two type of interactions also were considered. In the framework of general relativity it is accepted that a dark energy can explain the present cosmic acceleration. Except cosmological constant there are many others candidates of dark energy. The property of dark energy is model dependent and to differentiate different models of dark energy, a sensitive diagnostic tool is needed. Hubble parameter $H$ and deceleration parameter $q$ are very important quantities which can describe the geometric properties of the Universe. Since $\dot{a}>0$, hence $H>0$ means the expansion of the universe. Also, $\ddot{a}>0$, which is $q<0$ indicates the accelerated expansion of the Universe. Since, the various dark energy models give $H>0$ and $q<0$, they can not provide enough evidence to differentiate the more accurate cosmological observational data and the more general models of dark energy. For this aim we need  higher order of time derivative of scale factor and geometrical tool. Sahni \emph{et.al} \cite{Sahni} proposed geometrical statefinder diagnostic tool, based on dimensionless parameters $(r, s)$ which are function of scale factor and its time derivative. These parameters are defined as,
\begin{equation}\label{eq:statefinder}
r=\frac{1}{H^{3}}\frac{\dddot{a}}{a} ~~~~~~~~~~~~
s=\frac{r-1}{3(q-\frac{1}{2})}.
\end{equation}
In stellar astrophysics, the polytropic gas model can explain the
equation of state of degenerate white dwarfs, neutron stars and also the equation of state of
main sequence stars \cite{Christensen}. The idea of dark energy with polytropic gas equation of state has
been investigated by Mukhopadhyay and Ray in cosmology \cite{Mukhopadhyay}. In addition to statefinder diagnostic, the other analysis to discriminate between dark energy
models is $\omega-\omega'$ analysis that have been used widely in the papers \cite{Khodam}-\cite{Huang1}. Subject of our interest is to consider two different models and study cosmological parameters.\\
As we know the viscous cosmology is an important theory to describe the evolution of the
Universe. It means that the presence of viscosity in the fluid introduces many interesting
pictures in the dynamics of homogeneous cosmological models, which is used to study the
evolution of universe.\\
We consider composed models of a fluid consists of barotropic fluid $P=\omega \rho$ coupled with,
\begin{enumerate}
\item Viscous modified Chaplygin gas\\
\begin{equation}
P_{VCG}=A\rho_{\small{CG}}-\frac{B}{\rho_{\small{CG}}^{\alpha}}-3\xi H,
\end{equation}
\item Viscous Polytropic gas
\begin{equation}
P_{VPG}=K\rho_{PG}^{1+\frac{1}{n}}-3\xi H,
\end{equation}
\end{enumerate}
where $K$ and $n$ are the polytropic constant and polytropic index, respectively. The polytropic gas is a
phenomenological model of dark energy. The polytropic gas model has a type $III$, where the singularity
takes place at a characteristic scale factor $a_{s}$. Karami et al. investigated the interaction between dark energy and dark matter in polytropic gas scenario, the phantom behavior of polytropic gas, reconstruction of $f(T)$-gravity from the polytropic gas and the correspondence between polytropic gas and agegraphic dark energy model \cite{Karami}-\cite{Karami_2}. The cosmological implications of polytropic gas
dark energy model is also discussed in \cite{Malekjani_1}. The evolution of deceleration parameter in the
context of polytropic gas dark energy model represents the decelerated expansion at the
early universe and accelerated phase later as expected. The polytropic gas model has also
been studied from the viewpoint of statefinder analysis in \cite{Malekjani_2}.\\
There are several theoretical models to describe dark energy. Among them the model based on Chaplygin gas EoS and its extensions are interesting because of possibility of dynamical analysis and solving some famous problems in cosmological constant model.
Therefore, in order to construct a real model of our universe we consider the modified Chaplygin (or Polytropic) gas-like dark energy including viscosity and time-dependent interaction between components.
Above points are strong theoretical motivation to consider a toy model of our universe which needs observational data for confirmation or rejection.\\
This paper organized as follows. in the next section we will introduce the equations which governs our model. Then, we give numerical results corresponding both models. In the discussion section we summary our results. In two appendix we analyze more quantities of both models.
\section{\large{The field equations and models}}
Field equations that govern our model of consideration are,
\begin{equation}\label{eq:Einstein eq}
R^{\mu\nu}-\frac{1}{2}g^{\mu\nu}R^{\alpha}_{\alpha}=T^{\mu\nu}.
\end{equation}
By using the
following FRW metric for a flat Universe,
\begin{equation}\label{s2}
ds^2=-dt^2+a(t)^2\left(dr^{2}+r^{2}d\Omega^{2}\right),
\end{equation}
field equations can be reduced to the following Friedmann equations,
\begin{equation}\label{eq: Fridmman vlambda}
H^{2}=\frac{\dot{a}^{2}}{a^{2}}=\frac{\rho}{3},
\end{equation}
\begin{equation}\label{eq:Freidmann2}
\dot{H}=-\frac{1}{2}(\rho+P),
\end{equation}
where $d\Omega^{2}=d\theta^{2}+\sin^{2}\theta d\phi^{2}$, and $a(t)$
represents the scale factor. The $\theta$ and $\phi$ parameters are
the usual azimuthal and polar angles of spherical coordinates, with
$0\leq\theta\leq\pi$ and $0\leq\phi<2\pi$. The coordinates ($t, r,
\theta, \phi$) are called co-moving coordinates. Also $\rho$ and $p$ are total energy density and pressure respectively.\\
Energy conservation $T^{;j}_{ij}=0$ reads as,
\begin{equation}\label{eq:Bianchi eq}
\dot{\rho}+3H(\rho+P)=0.
\end{equation}
In order to introduce an interaction between DE and DM, we should mathematically split the equation (20) into two following equations,
\begin{equation}\label{eq:inteqm}
\dot{\rho}_{DM}+3H(\rho_{DM}+P_{DM})=Q,
\end{equation}
and,
\begin{equation}\label{eq:inteqG}
\dot{\rho}_{DE}+3H(\rho_{DE}+P_{DE})=-Q.
\end{equation}
For the barotropic fluid with $P_{DM}=\omega\rho_{DM}$, the equation (\ref{eq:inteqm}) will take following form,
\begin{equation}
\dot{\rho}_{b}+3H(1+\omega-b_{0}-btH)\rho_{b}=3H(b_{0}+btH)\rho_{CG},
\end{equation}
where the index $b$ refers to DM and $CG$ refers to dark energy. Dynamics of energy densities of Chaplygin and Polytropic gases reads as,
\begin{enumerate}
\item
\begin{equation}
\dot{\rho}_{CG}+3H(1+A+b_{0}+btH)\rho_{CG}-\frac{3HB}{\rho_{CG}^{\alpha}}=-3h(b_{0}+btH)\rho_{b}+9H^{2}\xi,
\end{equation}
\item
\begin{equation}
\dot{\rho}_{PG}+3H(1+K\rho_{PG}^{\frac{1}{n}}+b_{0}+btH)\rho_{PG}=-3h(b_{0}+btH)\rho_{b}+9H^{2}\xi.
\end{equation}
\end{enumerate}
In the above equation, index PG refers to Polytropic gases which serves as dark energy.
Cosmological parameters of our interest are EoS parameters of each components $\omega_{i}=P_{i}/\rho_{i}$ (index i refers to CG or PG), EoS parameter of composed fluid,
$$\omega_{tot}=\frac{P_{b}+P_{i} }{\rho_{b}+\rho_{i}},$$
and deceleration parameter $q$, which can be written as,
\begin{equation}\label{eq:accchange}
q=\frac{1}{2}(1+3\frac{P}{\rho} ),
\end{equation}
where $P=P_{b}+P_{i}$ and $\rho=\rho_{b}+\rho_{i}$. Hereafter, index $i$ means $CG$ (modified Chaplygin gas which usually written as MCG and PG for each model.
\section{\large{Numerical Results and Cosmological parameters}}
\subsection{\large{Model 1}}
This model is based on differential equation (24) which yields to the following results.\\\\
\begin{figure}[h!]
 \begin{center}$
 \begin{array}{cccc}
\includegraphics[width=50 mm]{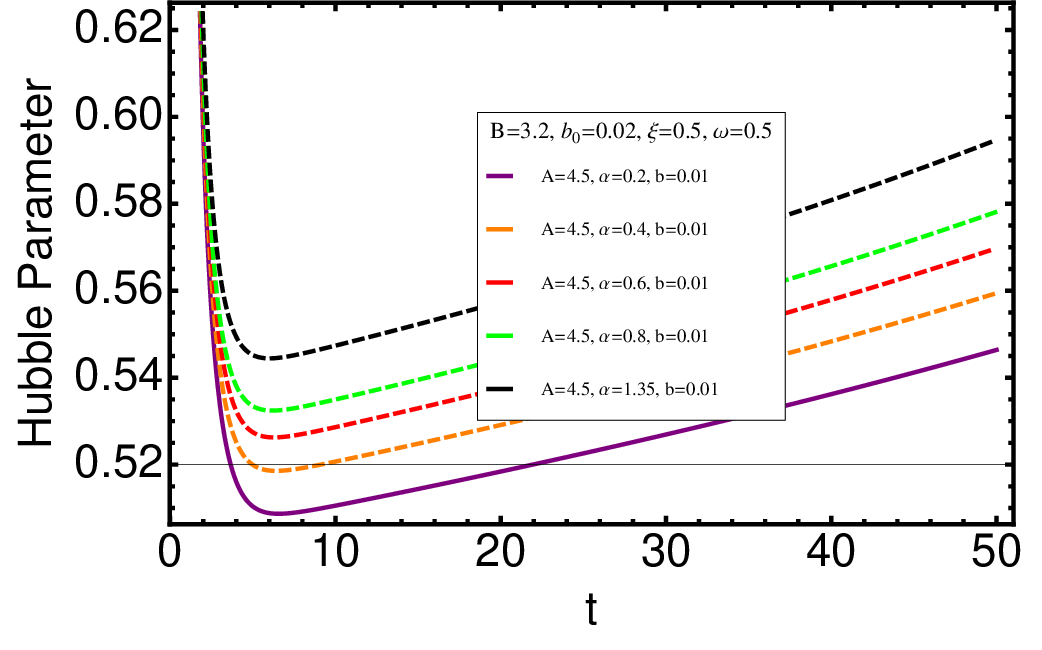} &
\includegraphics[width=50 mm]{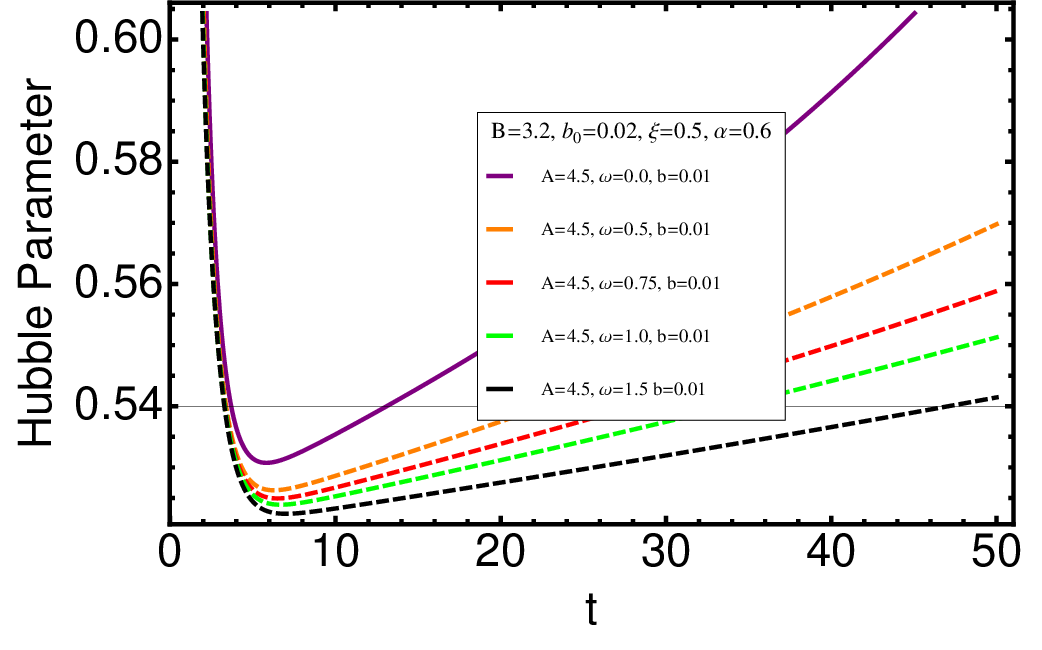}\\
\includegraphics[width=50 mm]{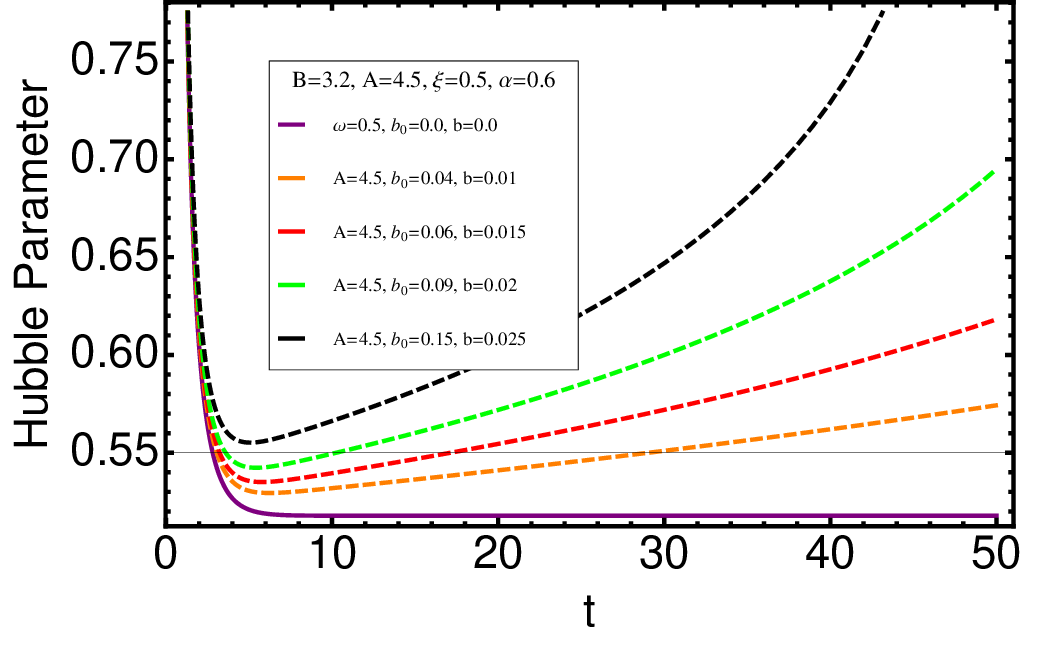} &
\includegraphics[width=50 mm]{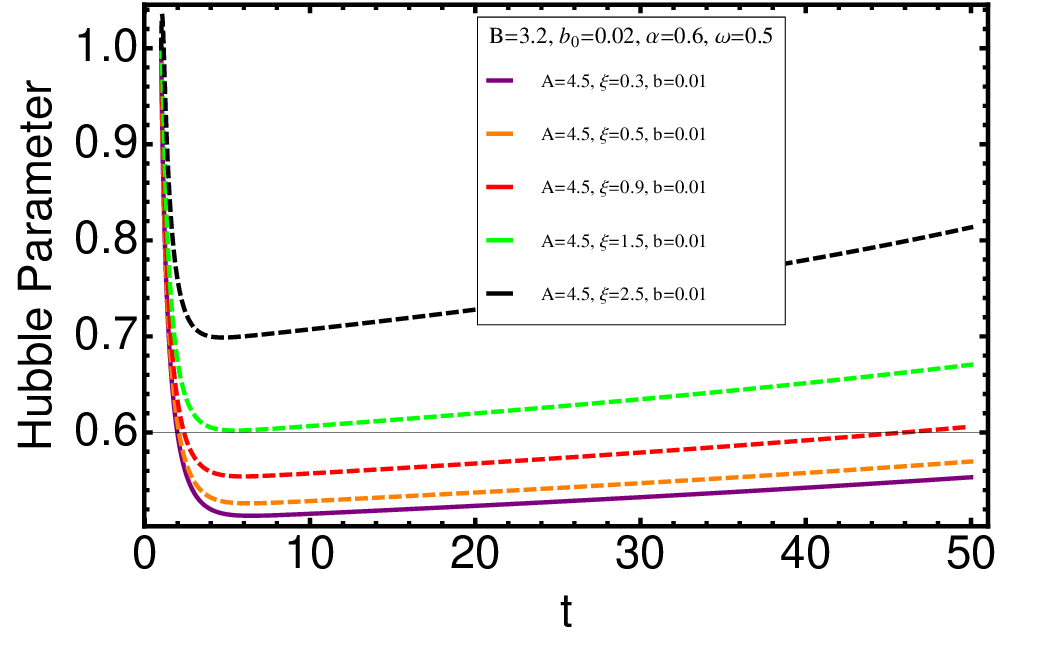}
 \end{array}$
 \end{center}
\caption{Behavior of Hubble parameter $H$ against $t$ for interacting barotropic fluid and viscous modified Chaplyagin gas.}
 \label{fig:1}
\end{figure}

Plots of the Fig. 1 show time evolution of Hubble expansion parameter in viscous modified Chaplygin gas model. In the first one we fixed all parameters and varies $\alpha$. We find that increasing $\alpha$ increases Hubble parameter. Also it is clear from the first plot of the Fig. 1 that evolution of Hubble parameter corresponding to low values of $\alpha$ is faster than higher values.\\
The second plot of the Fig. 1 shows behavior of Hubble expansion parameter with variation of $\omega$ which shows that increasing $\omega$ decreases Hubble parameter. Also it is clear from the second plot of the Fig. 1 (top right) that evolution of Hubble parameter corresponding to higher values of $\omega$ is faster than lower values.\\
In the next plot of the Fig. 1 (dawn left) we fixed all parameters and vary interaction parameters $b_{0}$ and $b$. We find that increasing interaction parameters increases Hubble parameter. Also it is clear that evolution of Hubble parameter corresponding to low values of interaction parameters is faster than higher values.\\
Finally the last plot of the Fig. 1 show the effect of viscosity. We find that increasing viscosity increases Hubble parameter. Also we find  that evolution of Hubble parameter corresponding to low values of viscosity is faster than higher values. This plot has more agreement with observational data which tells that $H_{0}\approx70$, where $H_{0}$ is current value of Hubble parameter which is corresponding to late time behavior of the figures. This behavior coincide with observational data for small value of the viscous parameter.\\
Plots of the Fig. 2 deal with time variation of total equation of state parameter. We see sudden evolution at initial stage, then  total equation of state parameter yields to approximately -1 as expected. We find from the first plot that increasing $\alpha$ decreases $\omega_{tot}$.\\
From the second plot We find that increasing $\omega$ increases $\omega_{tot}$. Also it is clear from the this plot that evolution of total equation of state parameter corresponding to low values of $\omega$ is faster than higher values.\\
In the third plot we can find variation of $\omega_{tot}$ with interaction parameters and find that these parameters decrease value of $\omega_{tot}$. We can see that in the case of without interaction ($b_{0}=b=0$) the value of total equation of state parameter take exactly -1 with condition $\omega_{tot}\geq-1$ which is quintessence like universe. Then, presence of interaction terms changed $\omega_{tot}$ to satisfy phantom like universe $\omega_{tot}\leq-1$.\\
Finally we find that viscous coefficient decrease value of $\omega_{tot}$. If we assume infinitesimal value of viscous parameter, then $\omega_{tot}\rightarrow-1$ verified with phantom regime [70].\\
Observational data needs to have $-1\leq\omega\leq-1/3$ which obtained by lower values of $b$ and $b_{0}$, or larger values of $\omega$ which illustrated in the second and third plots of the Fig. 2.\\
Plots of the Fig. 3 study behavior of $q$ against $t$ for interacting barotropic fluid and viscous modified Chaplyagin gas. We see similar behavior with the plots of the Fig. 2. Hence, we can say that $\alpha$, $b_{0}$, $b$ and $\xi$ decrease but $\omega$ increases value of the deceleration parameter. This case may agree with $\Lambda$CDM model (where $q\rightarrow-1$ observed) by choosing small values of interaction constants and larger value of $\omega$.\\

\begin{figure}[h!]
 \begin{center}$
 \begin{array}{cccc}
\includegraphics[width=50 mm]{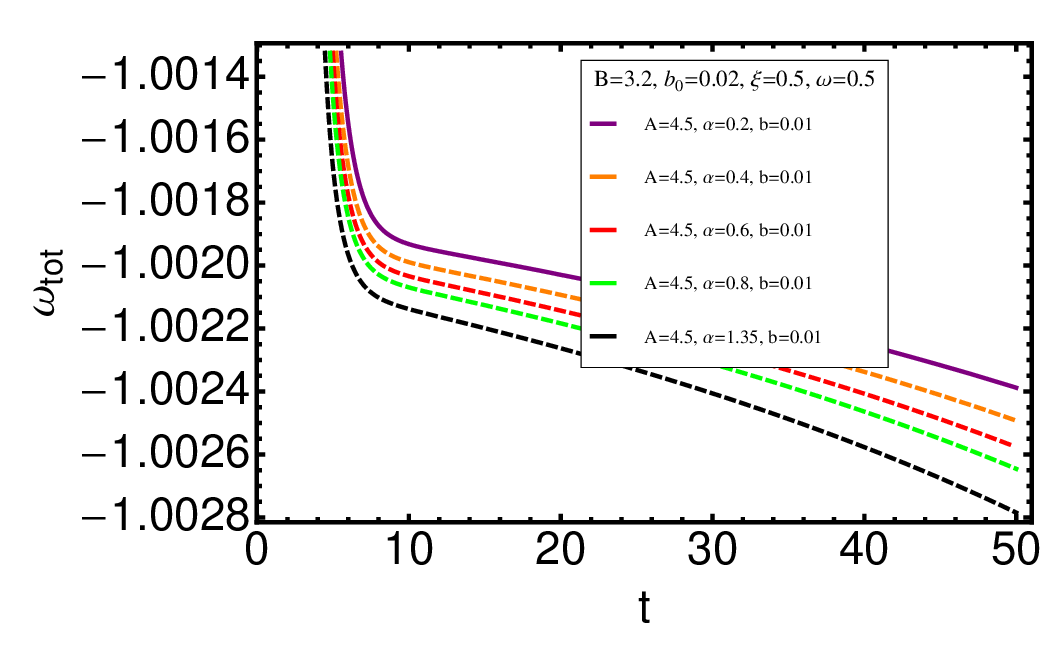} &
\includegraphics[width=50 mm]{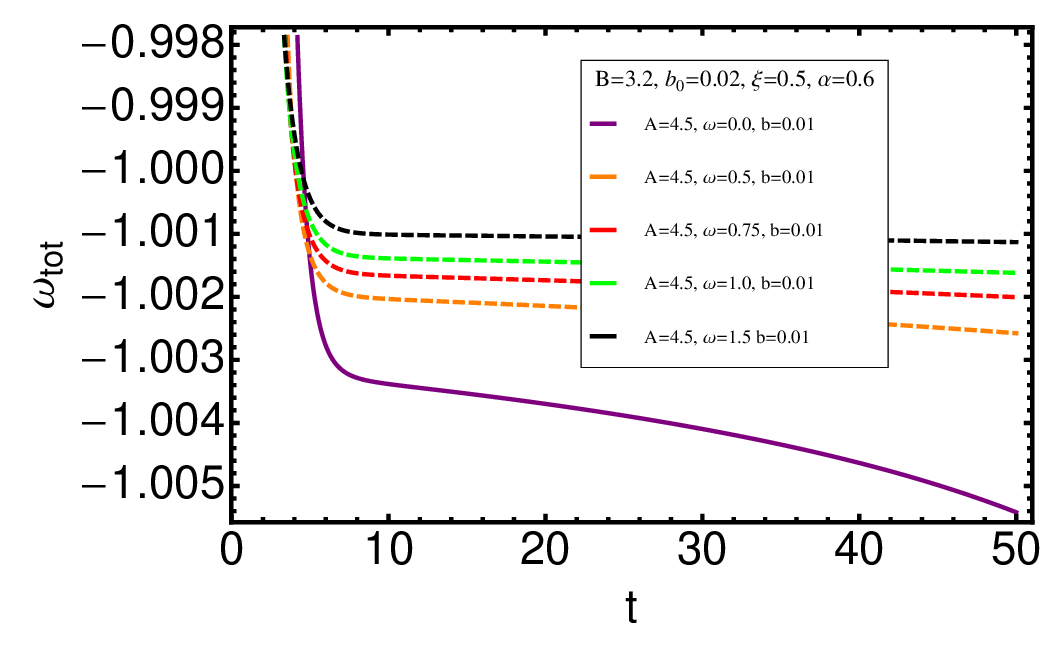}\\
\includegraphics[width=50 mm]{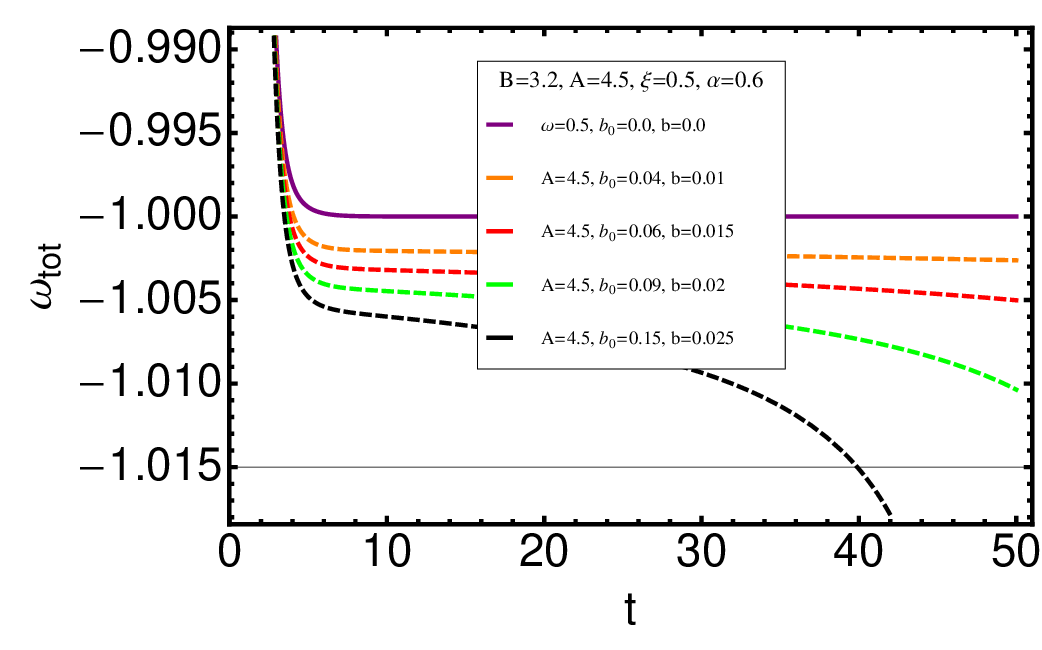} &
\includegraphics[width=50 mm]{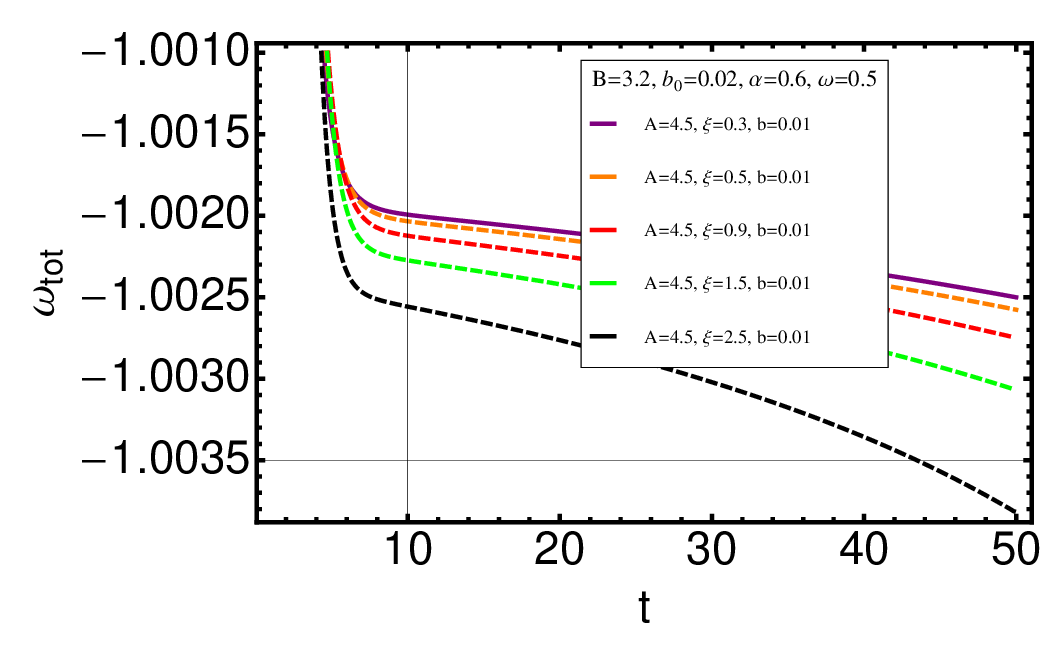}
 \end{array}$
 \end{center}
\caption{Behavior of EoS parameter $\omega_{tot}$ against $t$ for interacting barotropic fluid and viscous modified Chaplyagin gas.}
 \label{fig:2}
\end{figure}

\begin{figure}[h!]
 \begin{center}$
 \begin{array}{cccc}
\includegraphics[width=50 mm]{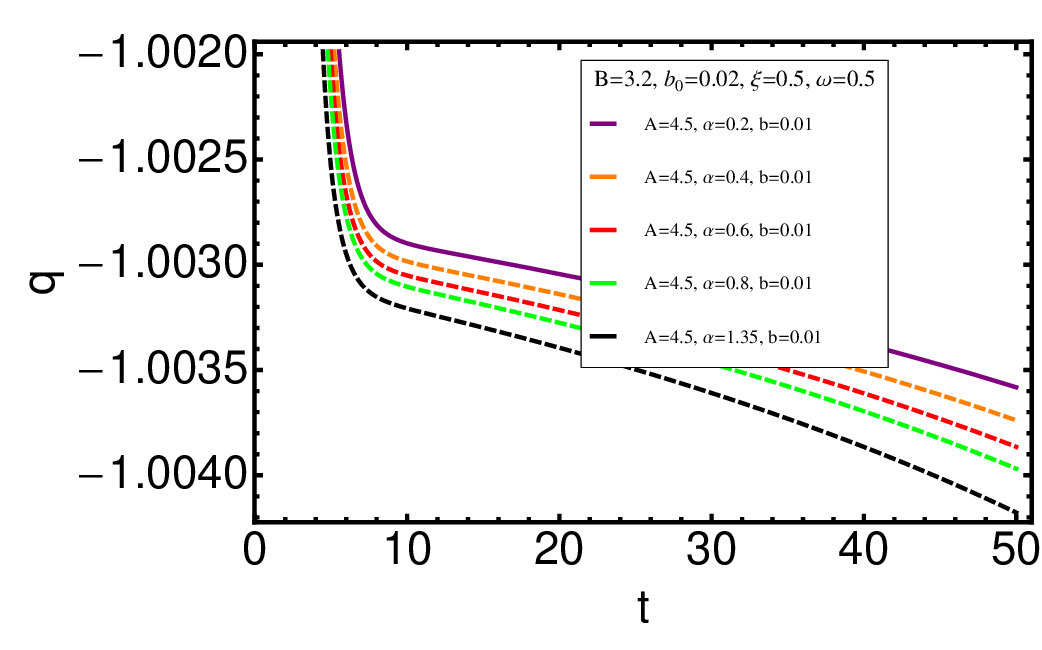} &
\includegraphics[width=50 mm]{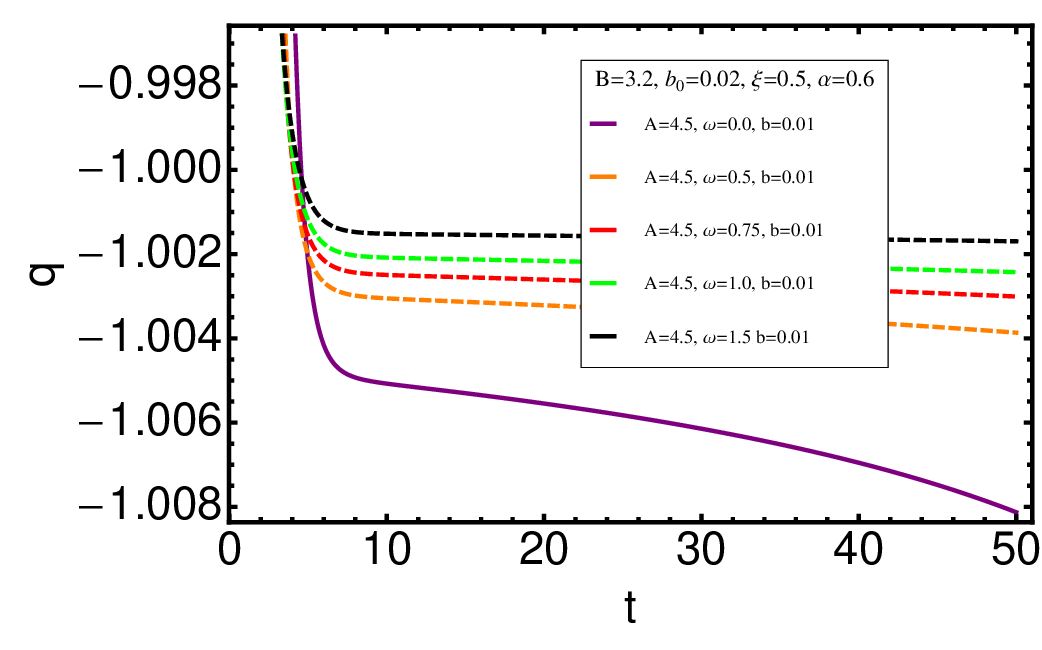}\\
\includegraphics[width=50 mm]{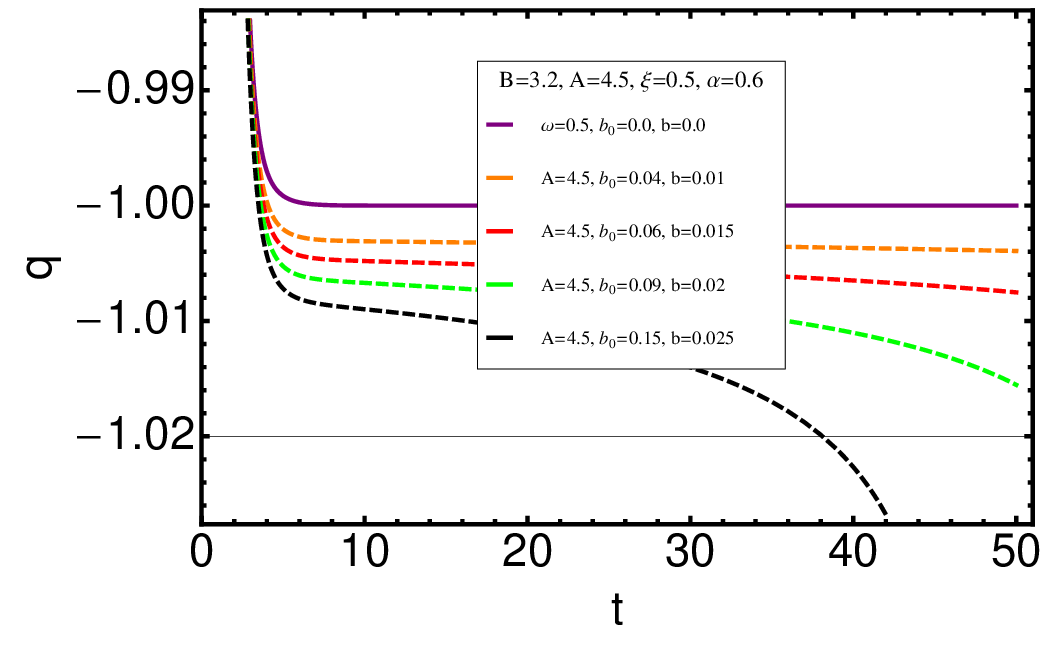} &
\includegraphics[width=50 mm]{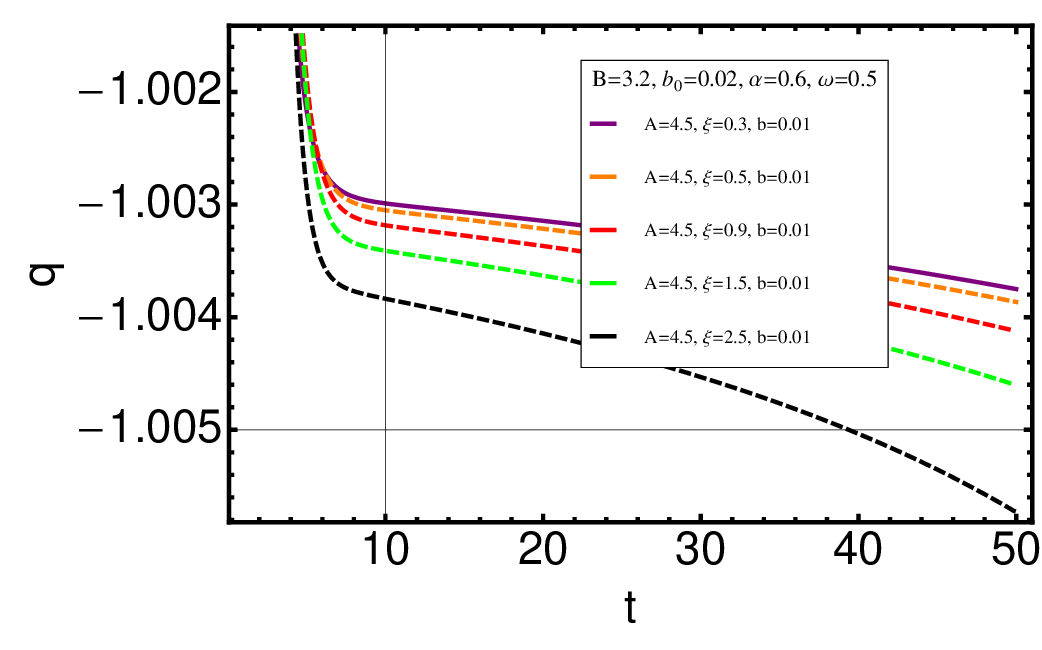}
 \end{array}$
 \end{center}
\caption{Behavior of deceleration parameter $q$ against $t$ for interacting barotropic fluid and viscous modified Chaplyagin gas.}
 \label{fig:3}
\end{figure}

\subsection{\large{Model 2}}
This model is based on differential equation (25) which yields to the following results.\\
Plots of the Fig. 4 show time evolution of Hubble expansion parameter in viscous polytropic gas model. We can see that the Hubble expansion parameter reduced suddenly at initial stage and take approximately constant value at the late time for appropriate parameters.\\
In the first plot of the Fig. 4 we fixed all parameters and varies $n$. We find that increasing $n$ decreases Hubble parameter. Also it is clear from the first plot of the Fig. 4 that evolution of Hubble parameter corresponding to higher values of $n$ is faster than lower values.\\
The second plot of the Fig. 4 shows behavior of Hubble expansion parameter with variation of $\omega$ which shows that increasing $\omega$ decreases Hubble parameter. Also it is clear from the second plot of the Fig. 4 that evolution of Hubble parameter corresponding to higher values of $\omega$ is faster than lower values. For the choice of $n=1.5$, $K=0.8$, $b_{0}=0.02$, $b=0.01$, $\xi=0.5$ and $\omega=0.5$ the Hubble expansion parameter yields to constant value at the late time.\\
In the next plot of the Fig. 4 we fixed all parameters and vary interaction parameters $b_{0}$ and $b$. We find that increasing interaction parameters decreases Hubble parameter which is opposite of the previous model. Also we can see that evolution of Hubble parameter corresponding to some values of interaction parameters is approximately similar.\\
Finally the last plot of the Fig. 4 show the effect of viscosity. We find that increasing viscosity increases Hubble parameter. Also, we find that evolution of Hubble parameter corresponding to low values of viscosity is faster than higher values.\\
It seems that the value of the viscosity in the interval $[0.5, 1.5]$ yields to more appropriate value of the current Hubble expansion parameter analogous to observational data.
Plots of the Fig. 5 deal with time variation of total equation of state parameter. We see sudden evolution at initial stage, then  total equation of state parameter yields to approximately -1 as expected, and similar to the previous model. We find from the first plot that increasing $n$ decreases $\omega_{tot}$ after suddenly evolution.\\
From the second plot We find that increasing $\omega$ increases $\omega_{tot}$. Also it is clear from the this plot that revolution of total equation of state parameter corresponding to low values of $\omega$ is faster than higher values which is similar to the previous model.\\
In the third plot we can find variation of $\omega_{tot}$ with interaction parameters and find that these parameters increase value of $\omega_{tot}$. We can see that in the case of without interaction ($b_{0}=b=0$) the value of total equation of state parameter takes closest value to -1. Then, presence of interaction terms changed $\omega_{tot}$ so we have $\omega_{tot}\leq-1$ (phantom regime) in the case of interacting.\\
Finally we find that viscous coefficient decreases value of $\omega_{tot}$. So, in this case, presence of viscosity is necessary to have $\omega_{tot}\rightarrow-1$.\\
Comparing with observational data suggests $\omega=0.75$ and $\xi=0.5$ are the best fitted values together with small values of interaction constants.

\begin{figure}[h!]
 \begin{center}$
 \begin{array}{cccc}
\includegraphics[width=50 mm]{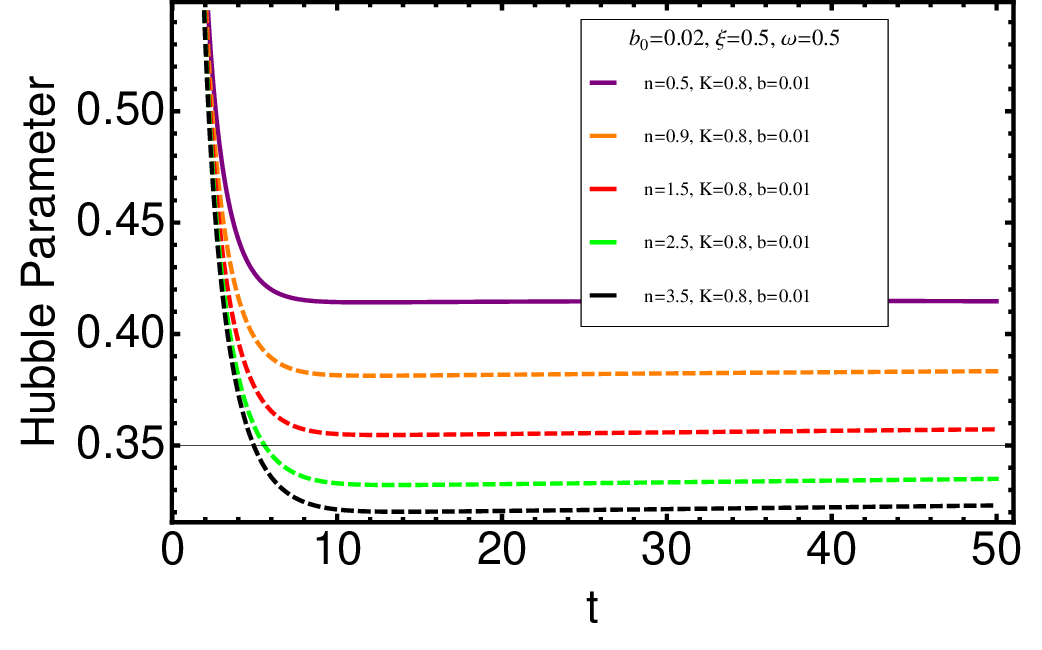} &
\includegraphics[width=50 mm]{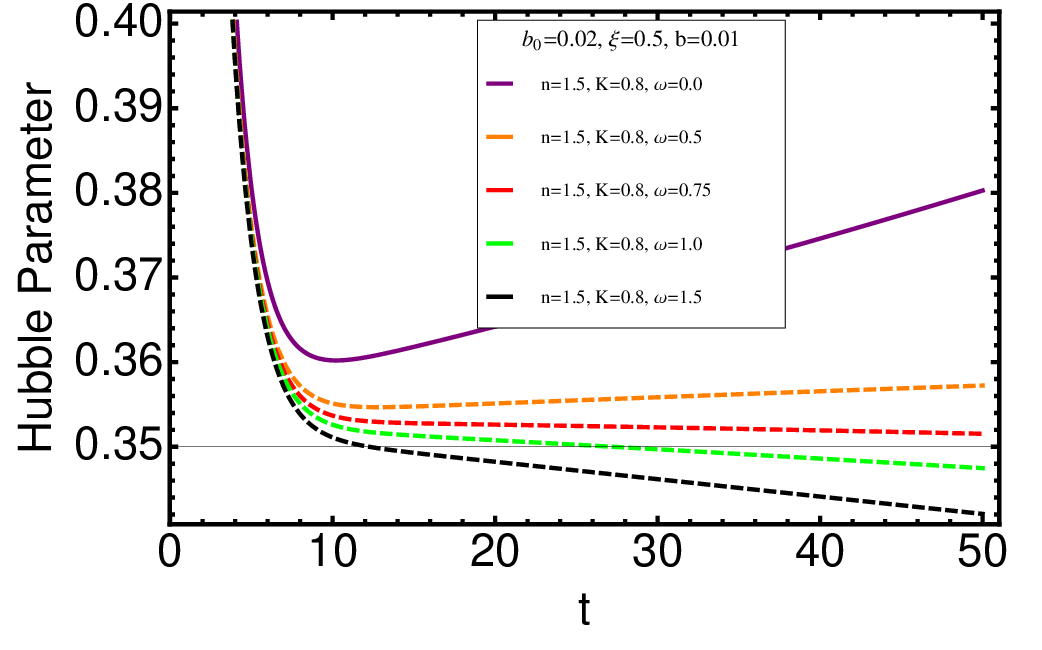}\\
\includegraphics[width=50 mm]{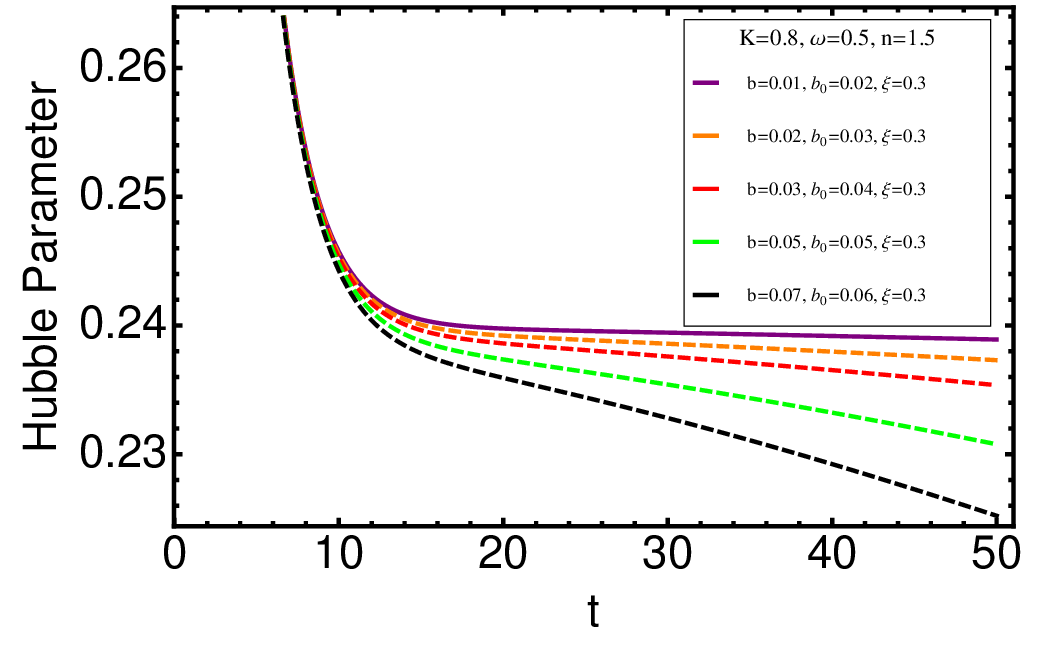} &
\includegraphics[width=50 mm]{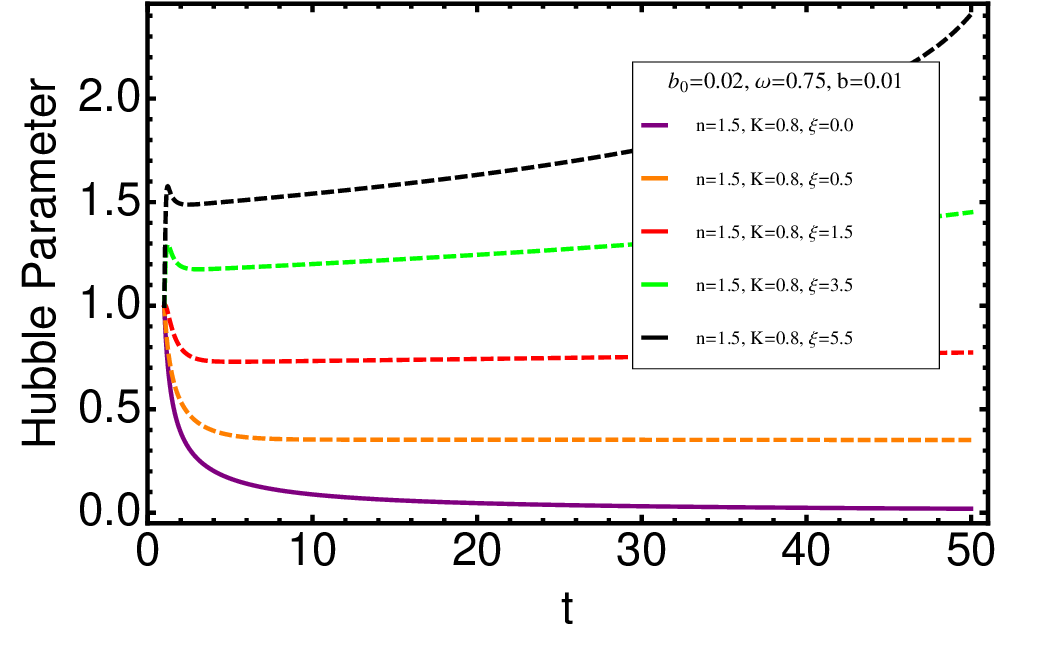}
 \end{array}$
 \end{center}
\caption{Behavior of Hubble parameter $H$ against $t$ for interacting barotropic fluid and viscous palotropic gas.}
 \label{fig:4}
\end{figure}

\begin{figure}[h!]
 \begin{center}$
 \begin{array}{cccc}
\includegraphics[width=50 mm]{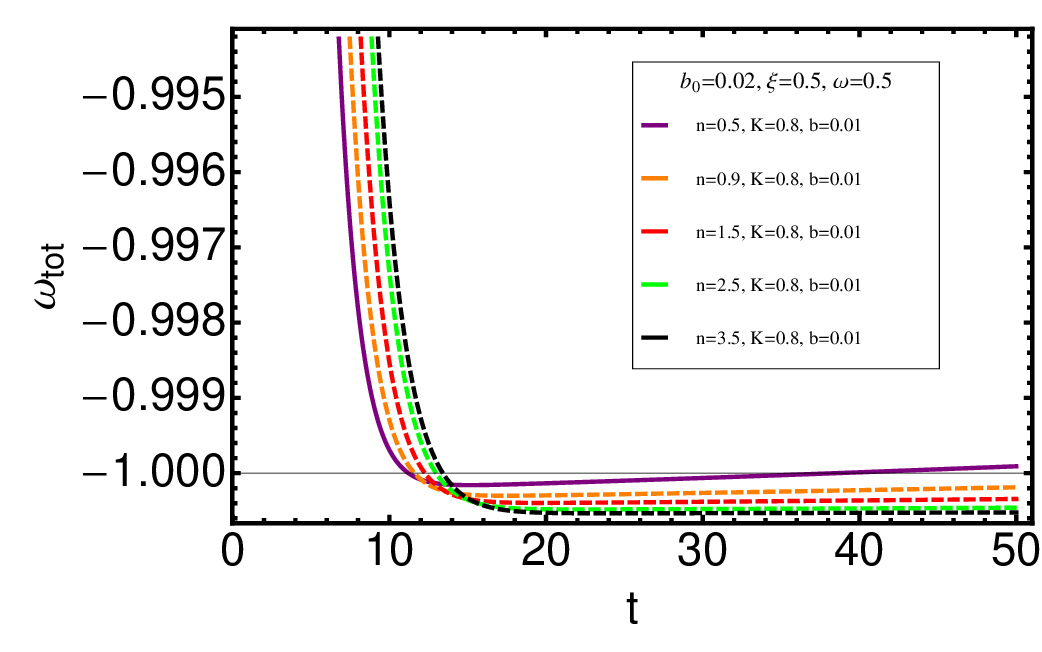} &
\includegraphics[width=50 mm]{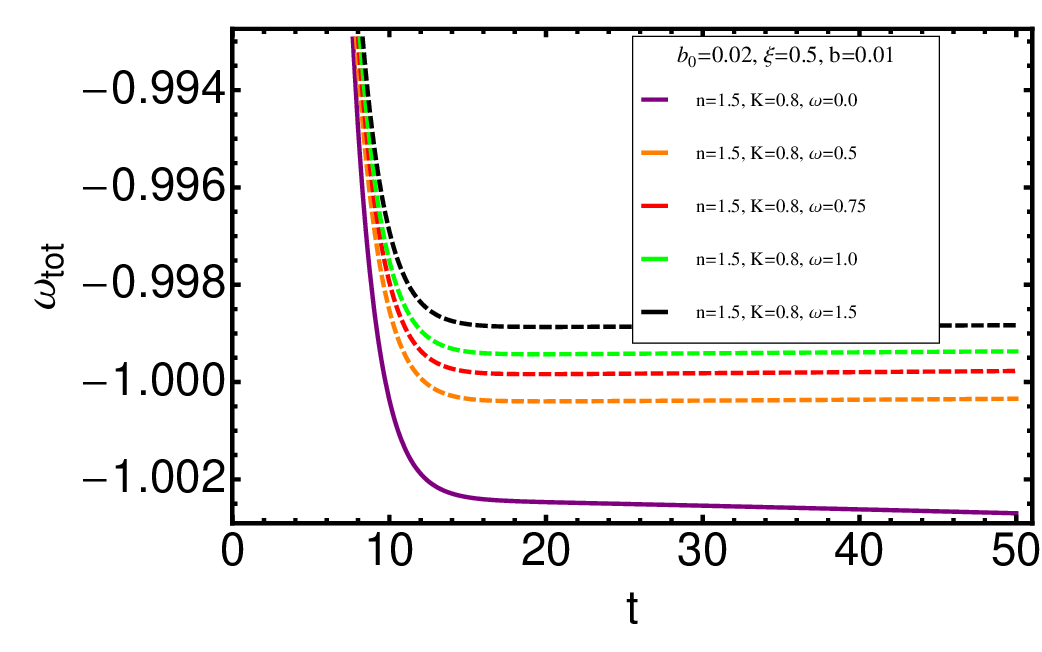}\\
\includegraphics[width=50 mm]{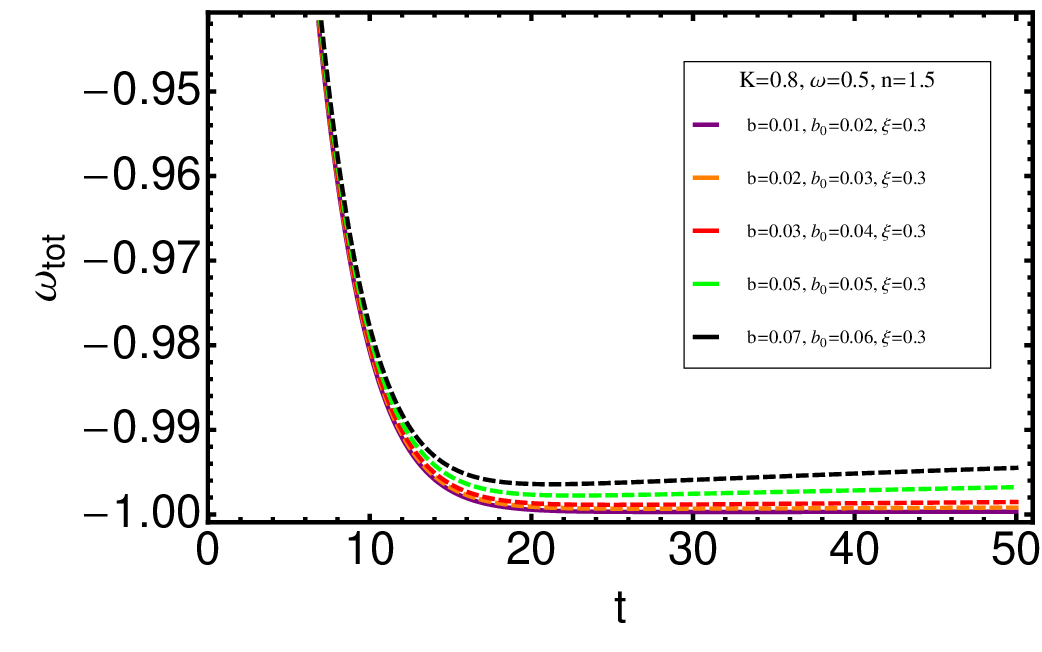} &
\includegraphics[width=50 mm]{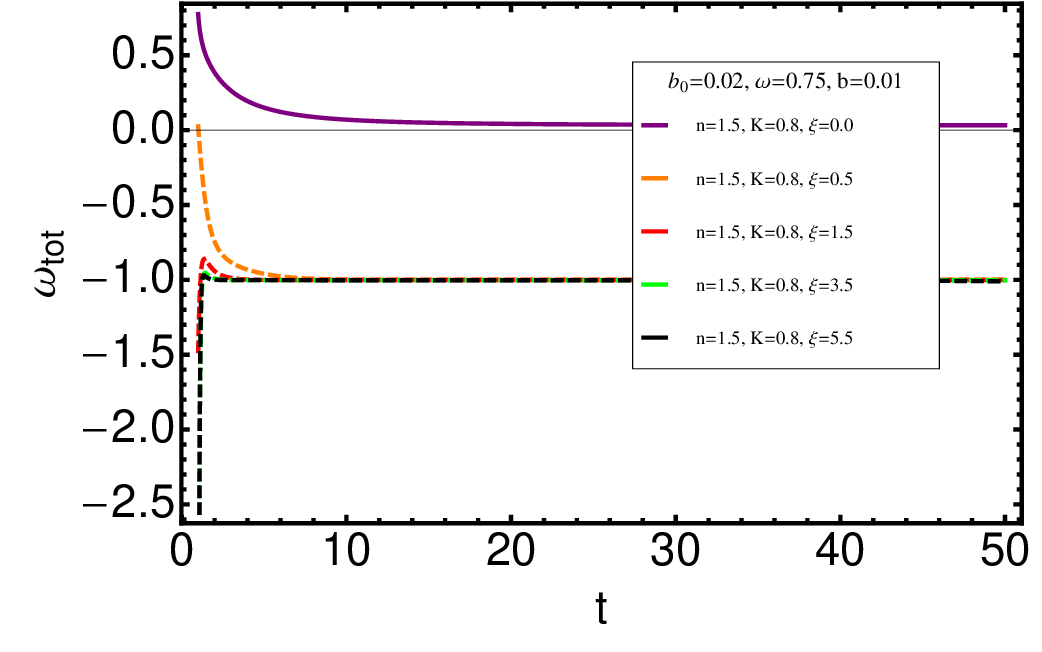}
 \end{array}$
 \end{center}
\caption{Behavior of EoS parameter $\omega_{tot}$ against $t$ for interacting barotropic fluid and viscous palotropic gas.}
 \label{fig:5}
\end{figure}

Plots of the Fig. 6 study behavior of deceleration parameter $q$ against $t$ for interacting barotropic fluid and viscous polytropic gas. We see similar behavior with the plots of the Fig. 5. Therefore, we can say that $n$ and $\xi$ decrease but $\omega$, $b_{0}$ and $b$ increase value of the deceleration parameter. This model is also agree with $\Lambda$CDM where $q\rightarrow-1$.\\
In the appendix A and B we study behavior of further cosmological parameters of both models such as energy density and pressure.

\begin{figure}[h!]
 \begin{center}$
 \begin{array}{cccc}
\includegraphics[width=50 mm]{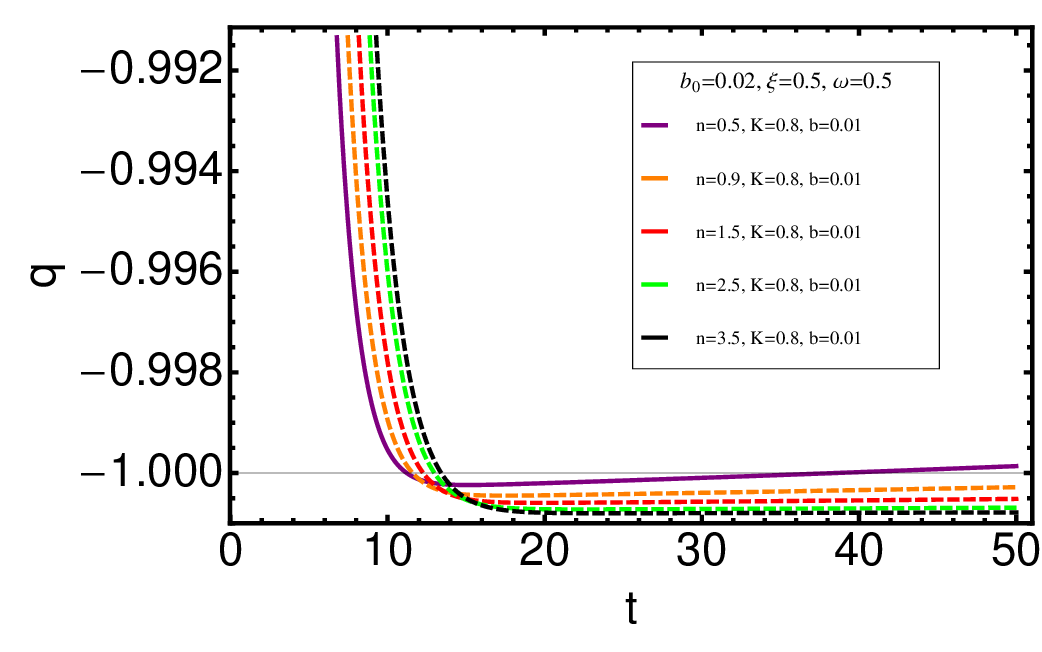} &
\includegraphics[width=50 mm]{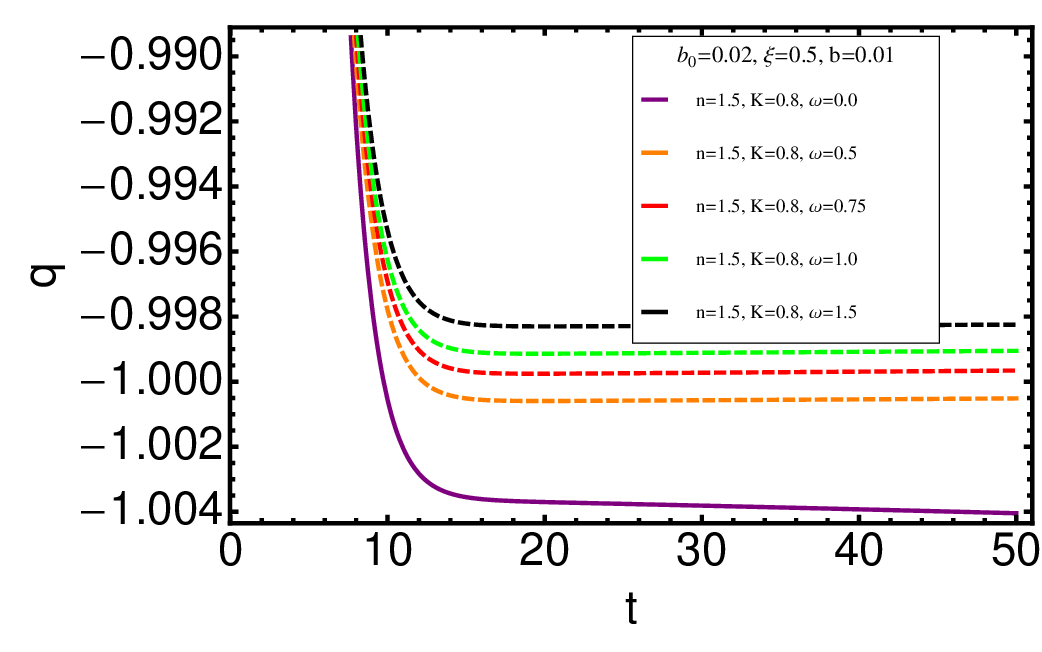}\\
\includegraphics[width=50 mm]{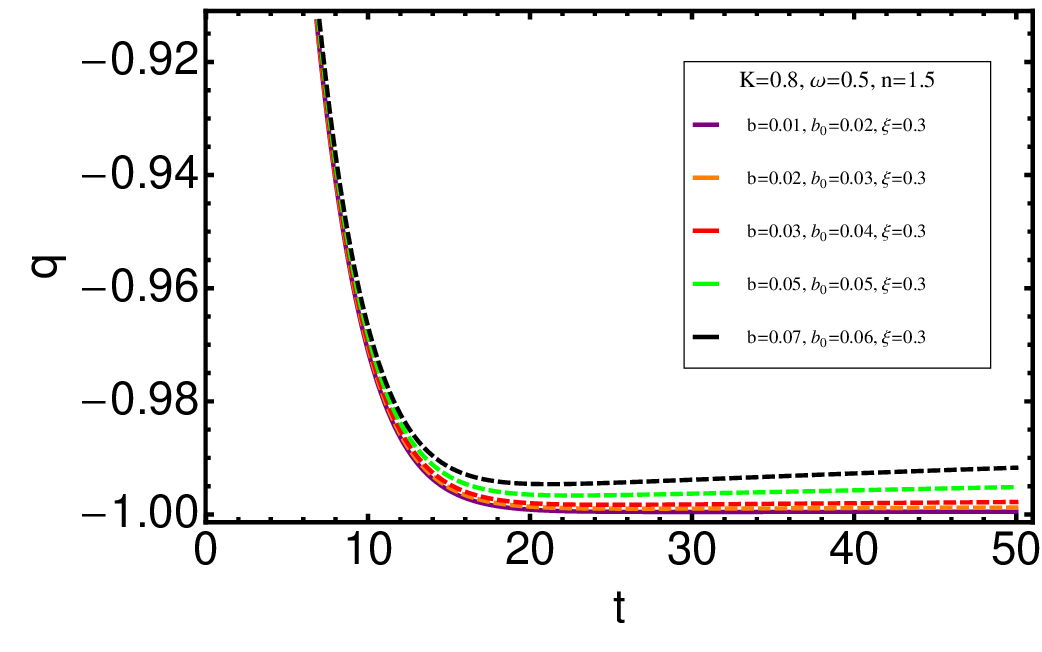} &
\includegraphics[width=50 mm]{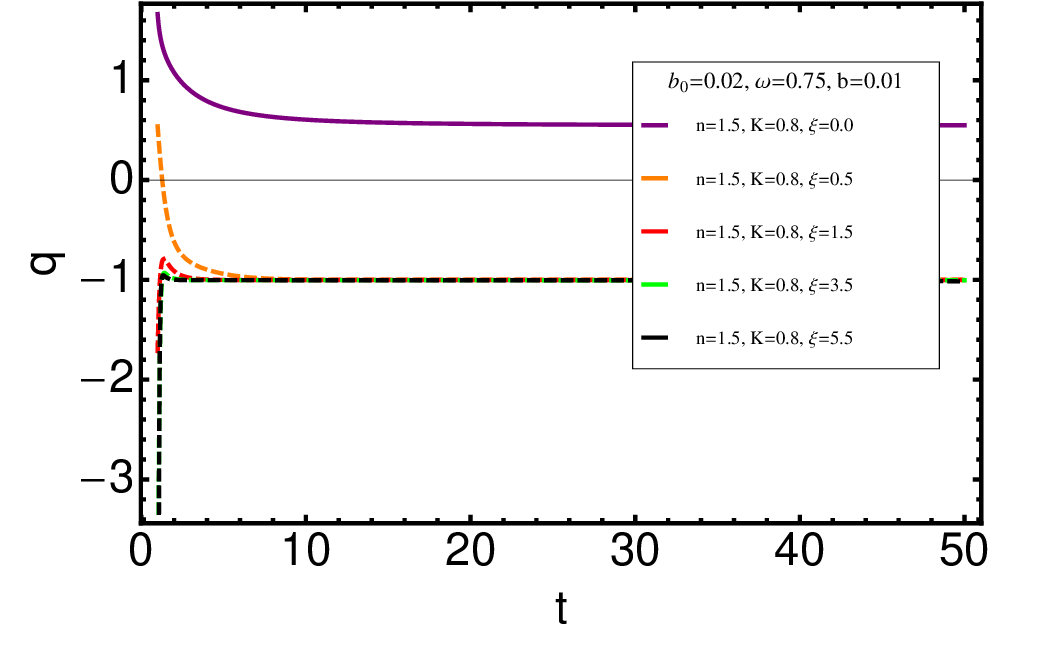}
 \end{array}$
 \end{center}
\caption{Behavior of deceleration parameter $q$ against $t$ for interacting barotropic fluid and viscous palotropic gas.}
 \label{fig:6}
\end{figure}

\section{Discussion}
We considered two different models of viscous interacting cosmology with modified interaction term so it is depend on Hubble parameter and discussed numerically cosmological parameters of the models. In the first model we consider viscous modified Chaplygin gas which interact with barotropic fluid. We obtained effect of interaction and viscous parameters on the cosmological quantities. We found that these parameters increase Hubble expansion parameter. If we neglect interaction parameters and viscosity, then evolution of Hubble parameter is faster than the case of interacting viscous cosmology. In the non-interacting case the Hubble parameter yields to constant after sudden reduction at initial stage. Also we studied equation of state parameters and found that interaction parameters and viscosity decrease value of EoS parameters. This situation is similar for deceleration parameter. In the non-interacting case, EoS and deceleration parameters yields to -1 as expected. We, then studied effect of these parameters on total density and pressure. We found that both interaction parameters and viscosity increase value of total density but decrease value of total pressure. At the initial stage the total density suddenly decreased and yields to a constant for non-interacting case, but it is increasing function of time in presence of interaction term. We show that this model may agree with some observational data which tell $H_{0}\approx70$ (in our scale $H_{0}\approx0.7$), and $q\rightarrow-1$. \\
In the second model we consider viscous polytropic gas which interact with barotropic fluid. Just before, we obtained effect of interaction and viscous parameters on the cosmological quantities. We found that interaction parameters decrease but viscosity increases Hubble expansion parameter. Behavior of interaction term in Hubble expansion parameter of this model is opposite of previous model. If we neglect interaction parameters and viscosity, then evolution of Hubble parameter is faster than the case of interacting viscous cosmology. In the non-interacting or non-viscous cases the Hubble parameter yields to approximately a constant after sudden reduction at initial stage. Also we studied equation of state parameters and found that interaction parameters increase and viscosity decreases value of EoS parameters. EoS parameter yields to -1 for the non-interacting case and yields to 0 for non-viscous case. The effect of interaction parameters on the deceleration parameter is similar to the EoS parameter but the deceleration parameter yields to approximately 0.5 for the non-viscous cosmology.  Finally we studied effect of these parameter on total density and pressure. We found that interaction parameters decrease but viscosity increases value of total density. On the other hand interaction parameters increase total pressure but viscosity decreases one. This model is also may agree with some observational data even more than the first model. In both models, the phantom regime obtained by adding interaction and we have $\omega_{tot}\leq-1$. However further studies such as [71] are needed to confirm
the viability of these models.\\
For the future work it is interesting to consider the effects of varying viscosity [72] on the cosmological parameters of present model.

\newpage
\section*{Appendix A}
In this appendix we study equation of state parameter corresponding to viscous modified Chaplygin gas, total density and pressure of the model numerically. Plots of the Fig. 7 show time evolution of $\omega_{VMCG}$ with variation of $\alpha$, $\omega$, $b_{0}$, $b$ and $\xi$. We find $\alpha$, $b_{0}$, $b$ and $\xi$ decrease value of equation of state parameter but $\omega$ increased one. Then, plots of the Fig. 8 show that total density increase by $\alpha$, $b_{0}$, $b$ and $\xi$, but pressure decrease with these parameters (see Fig. 9).
\begin{figure}[h!]
 \begin{center}$
 \begin{array}{cccc}
\includegraphics[width=50 mm]{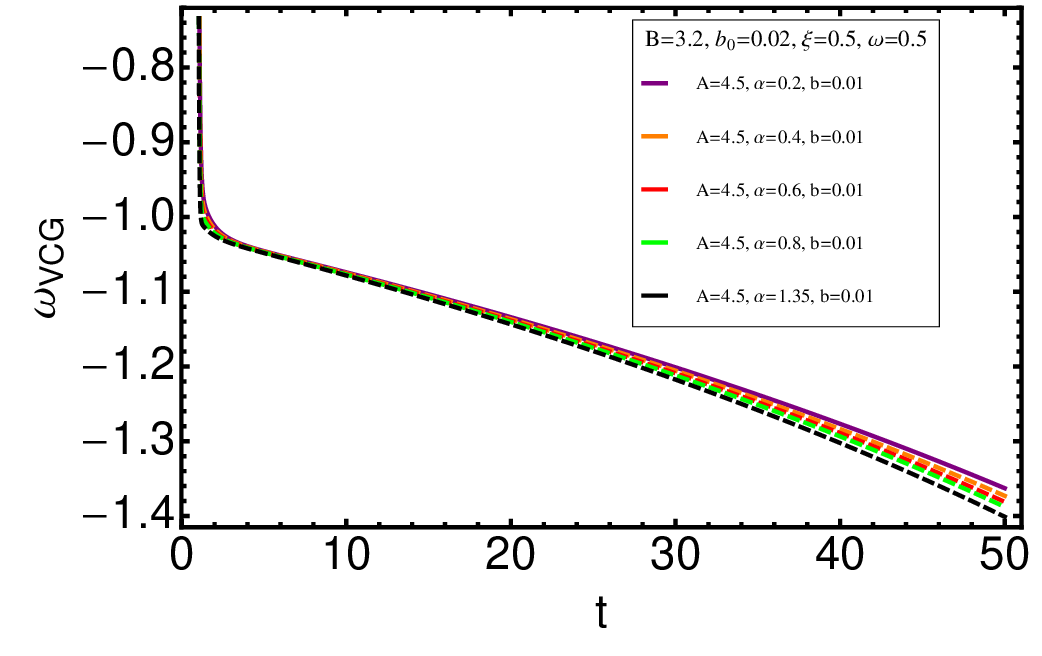} &
\includegraphics[width=50 mm]{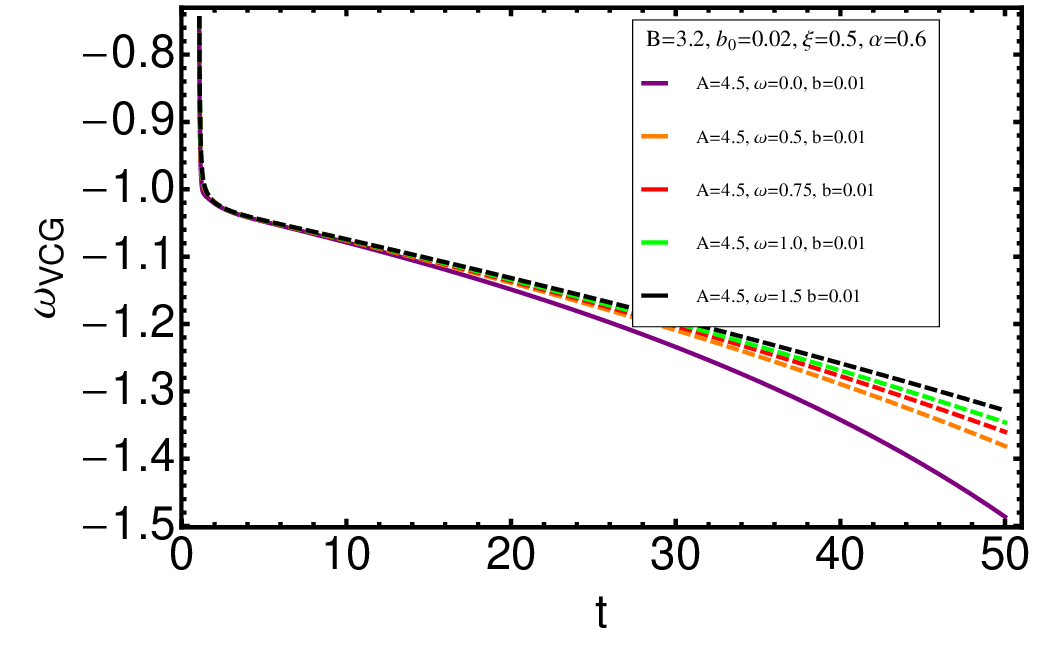}\\
\includegraphics[width=50 mm]{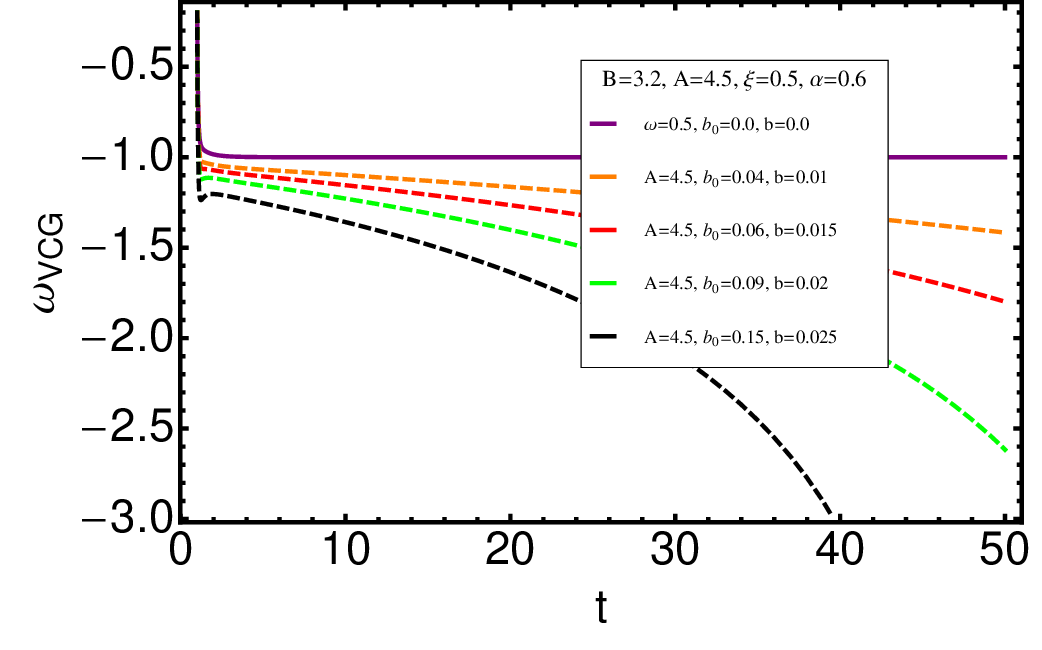} &
\includegraphics[width=50 mm]{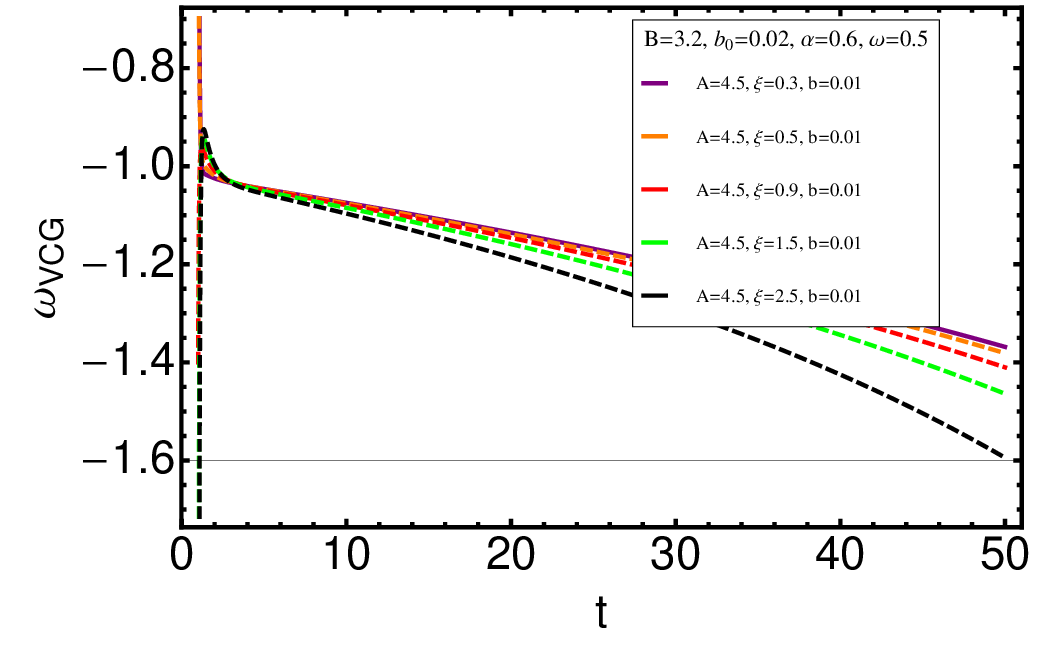}
 \end{array}$
 \end{center}
\caption{Behavior of EoS parameter of viscous Chaplyagin $\omega_{VCG}$ against $t$ for interacting barotropic fluid and viscous Chaplyagin gas.}
 \label{fig:7}
\end{figure}
\begin{figure}[h!]
 \begin{center}$
 \begin{array}{cccc}
\includegraphics[width=50 mm]{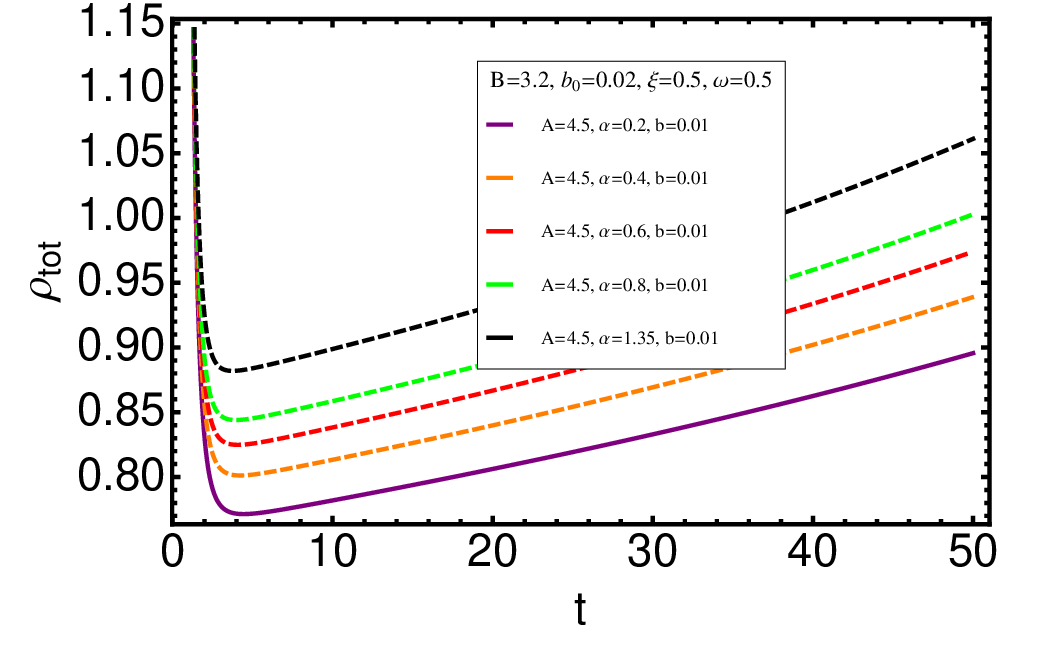} &
\includegraphics[width=50 mm]{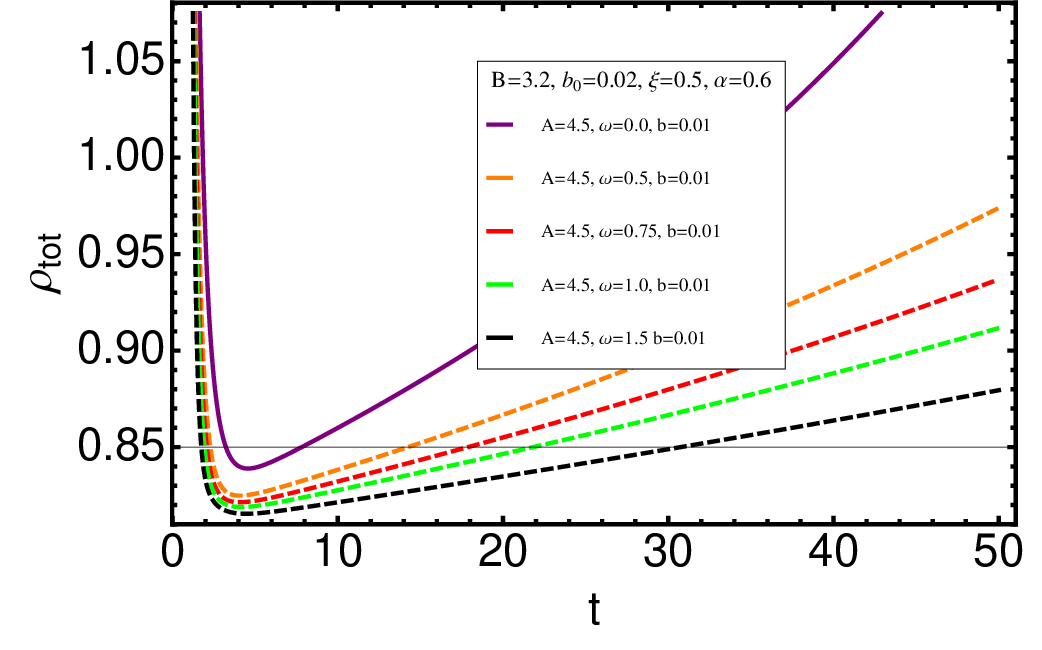}\\
\includegraphics[width=50 mm]{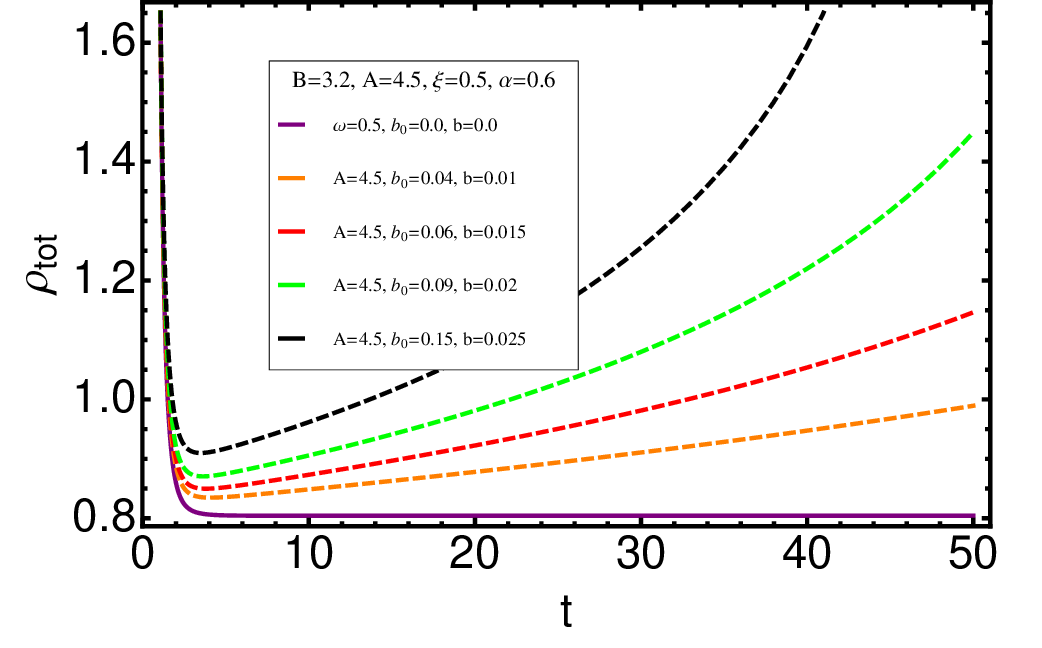} &
\includegraphics[width=50 mm]{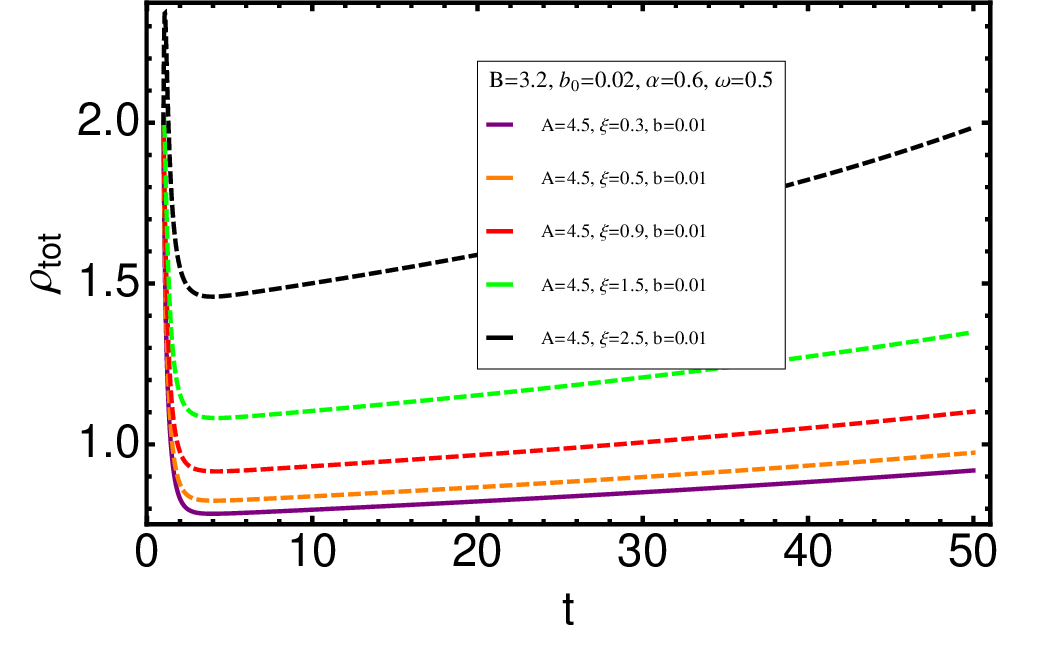}
 \end{array}$
 \end{center}
\caption{Behavior of $\rho_{tot}$ against $t$ for interacting barotropic fluid and viscous Chaplyagin gas.}
 \label{fig:8}
\end{figure}
\begin{figure}[h!]
 \begin{center}$
 \begin{array}{cccc}
\includegraphics[width=50 mm]{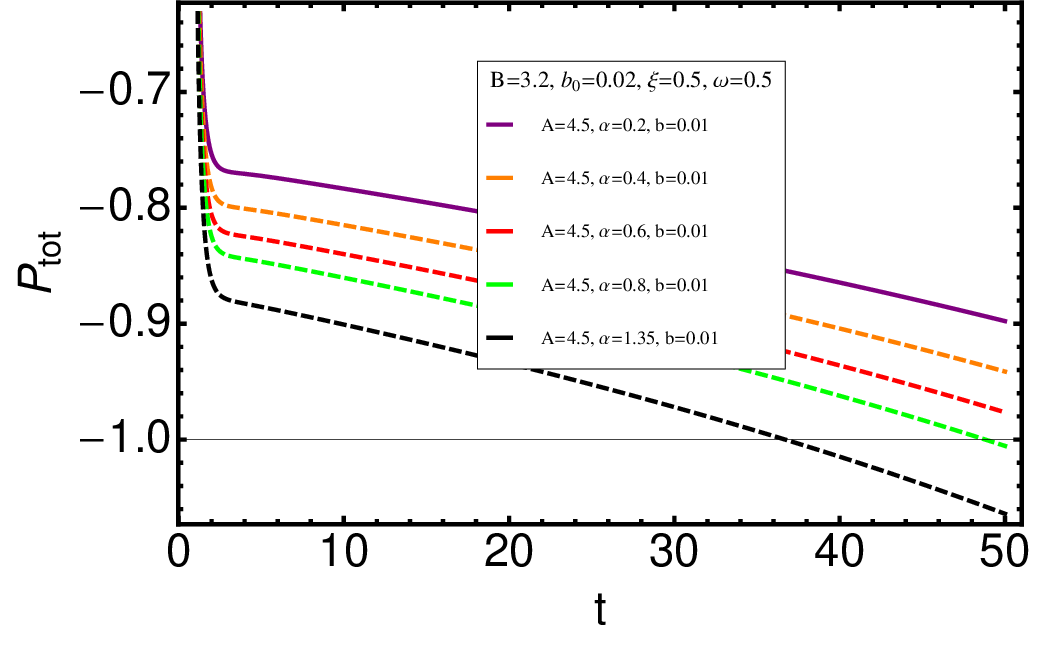} &
\includegraphics[width=50 mm]{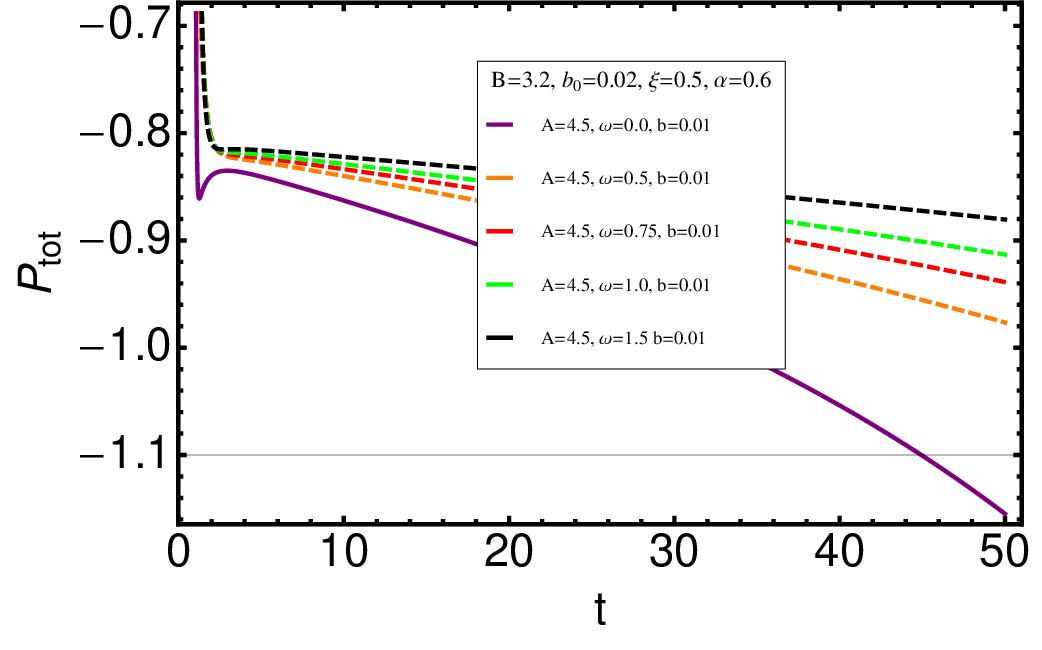}\\
\includegraphics[width=50 mm]{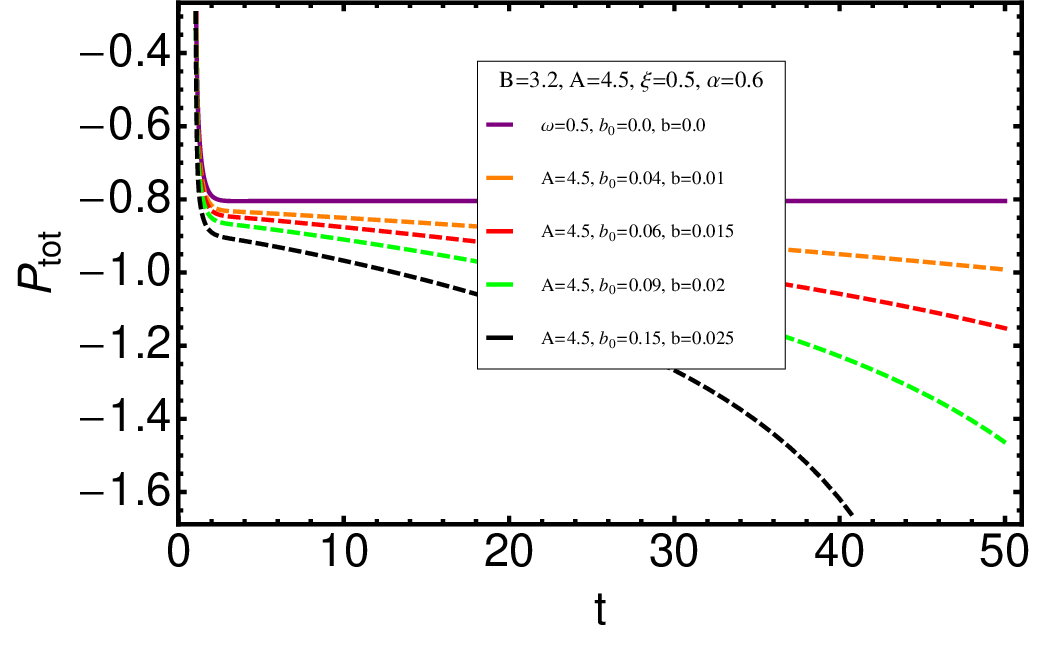} &
\includegraphics[width=50 mm]{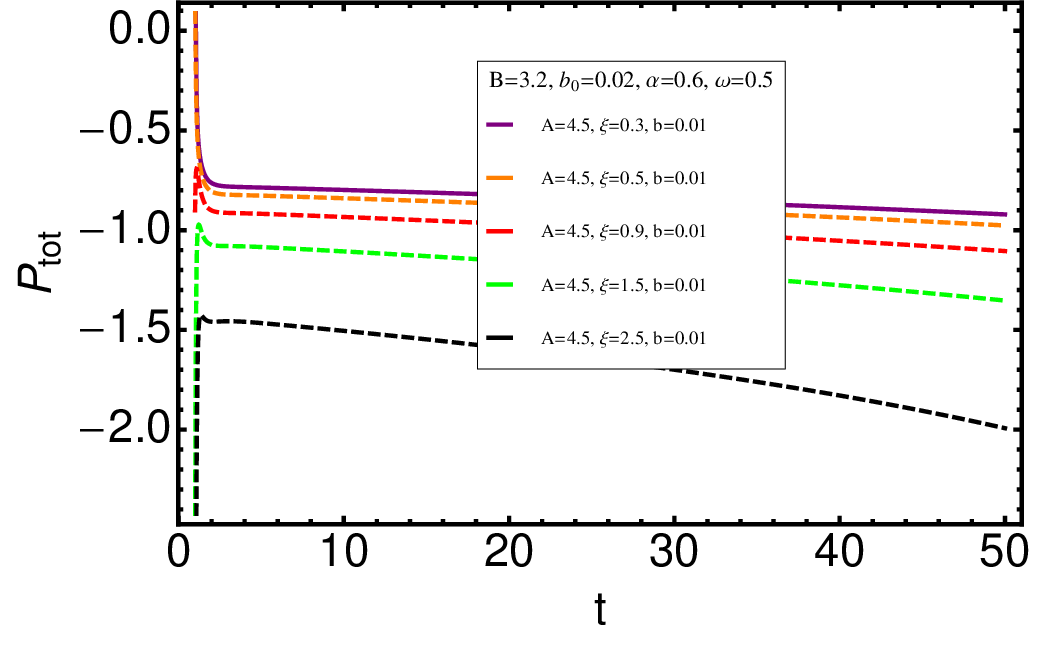}
 \end{array}$
 \end{center}
\caption{Behavior of $P_{tot}$ against $t$ for interacting barotropic fluid and viscous Chaplyagin gas.}
 \label{fig:9}
\end{figure}

\newpage
\section*{Appendix B}
In this appendix we study equation of state parameter corresponding to viscous palotropic gas, total density and pressure of the model numerically. Plots of the Fig. 10 show time evolution of $\omega_{VPG}$ with variation of $n$, $\omega$, $b_{0}$, $b$ and $\xi$. We find $n$ and $\omega$ increase value of equation of state parameter but $b_{0}$, $b$ and $\xi$ decreased one. Then, plots of the Fig. 11 show that total density increase by $\xi$ and decrease by $b_{0}$, $b$ and $n$, but total pressure decrease with $\xi$ and increase with other parameters (see Fig. 12).

\begin{figure}[h!]
 \begin{center}$
 \begin{array}{cccc}
\includegraphics[width=50 mm]{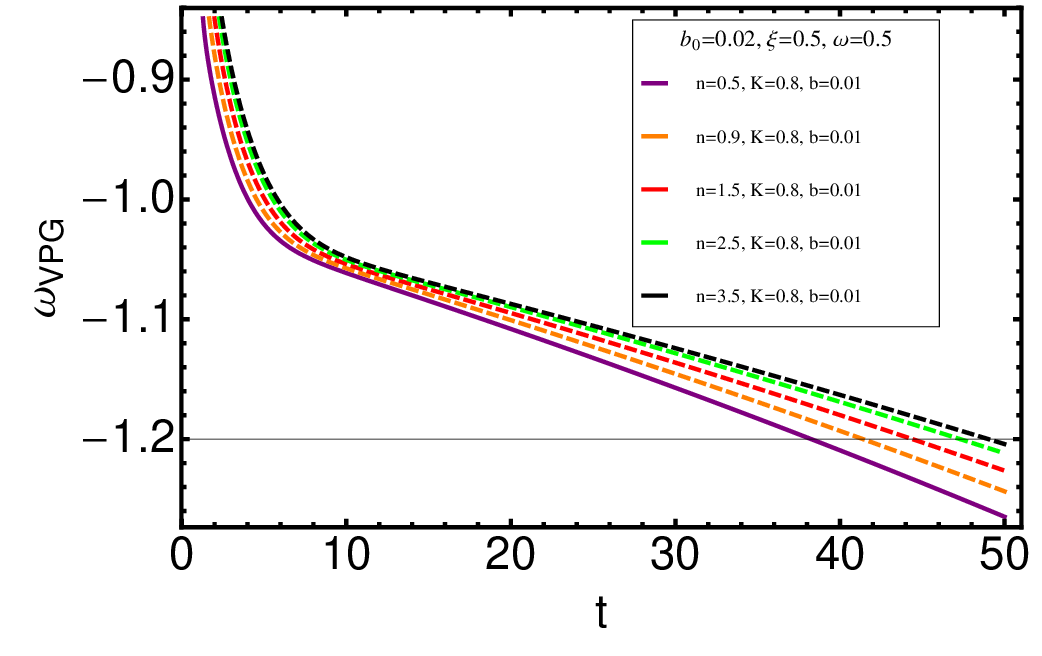} &
\includegraphics[width=50 mm]{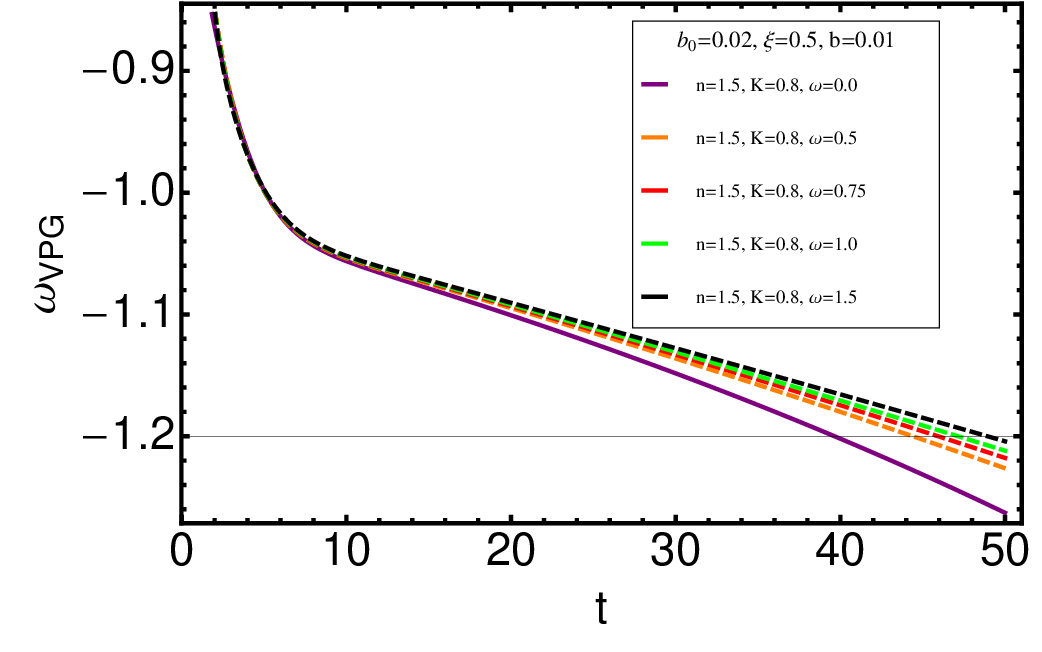}\\
\includegraphics[width=50 mm]{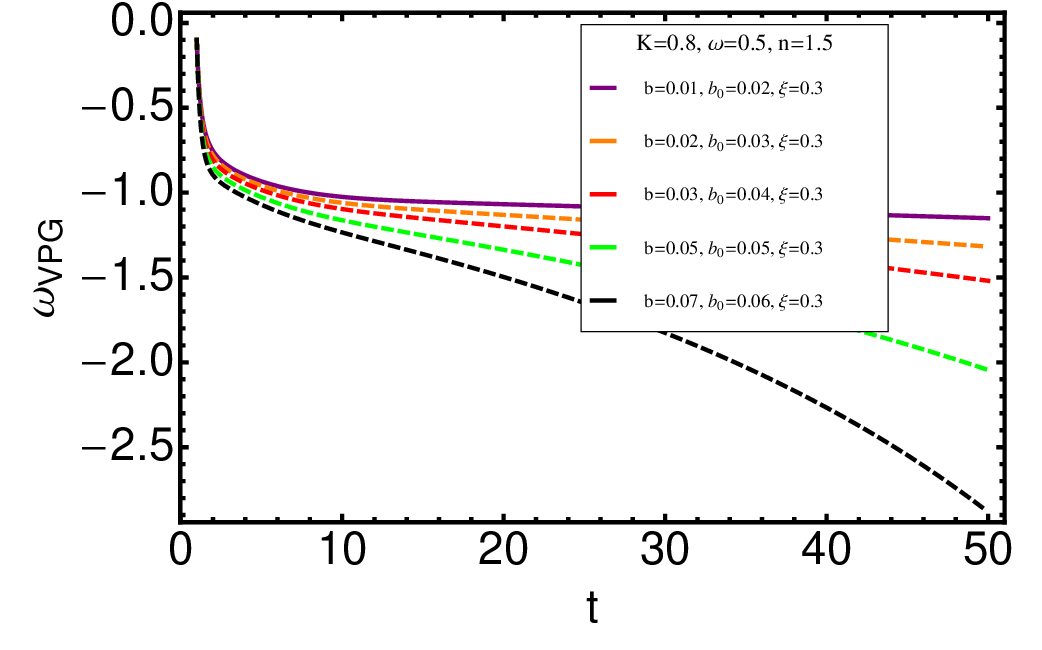} &
\includegraphics[width=50 mm]{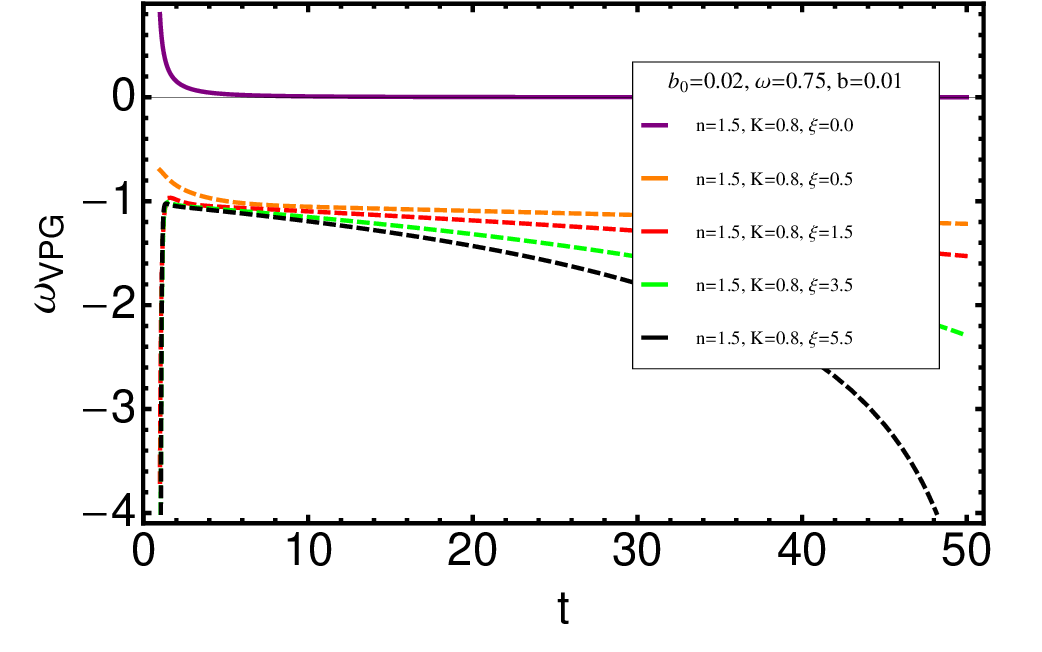}
 \end{array}$
 \end{center}
\caption{Behavior of EoS parameter of viscous palotropic $\omega_{VCG}$ against $t$ for interacting barotropic fluid and viscous palotropic gas.}
 \label{fig:10}
\end{figure}

\begin{figure}[h!]
 \begin{center}$
 \begin{array}{cccc}
\includegraphics[width=50 mm]{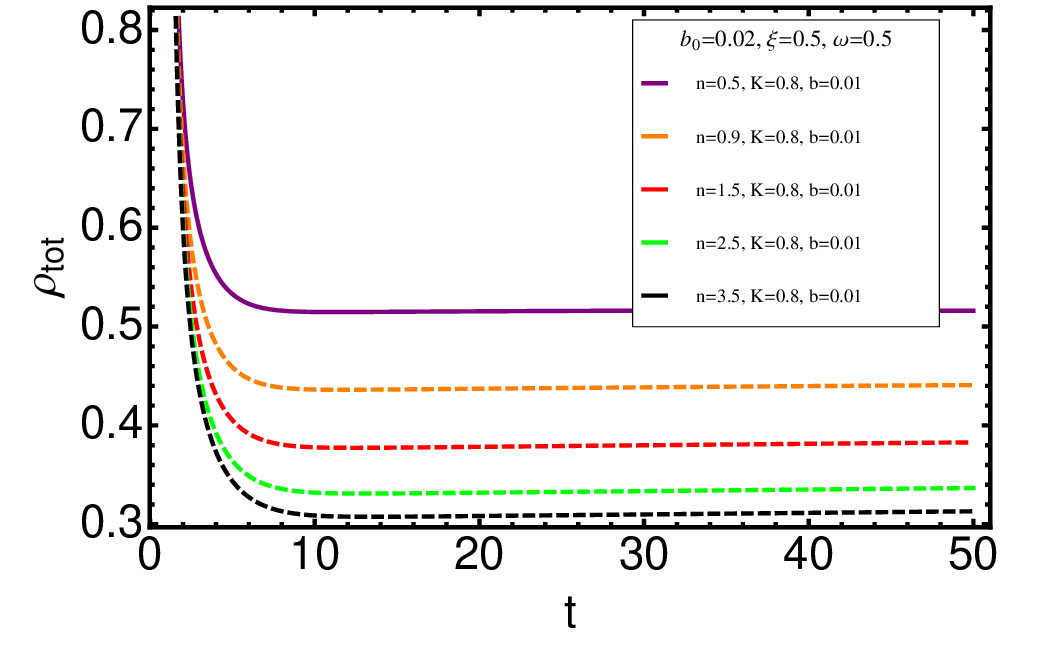} &
\includegraphics[width=50 mm]{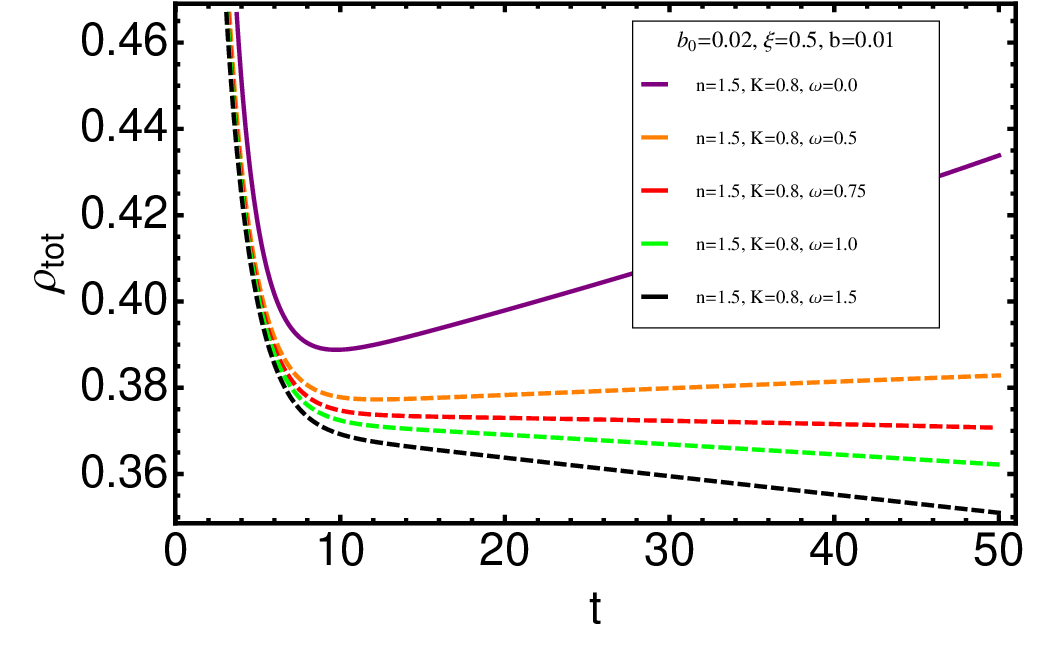}\\
\includegraphics[width=50 mm]{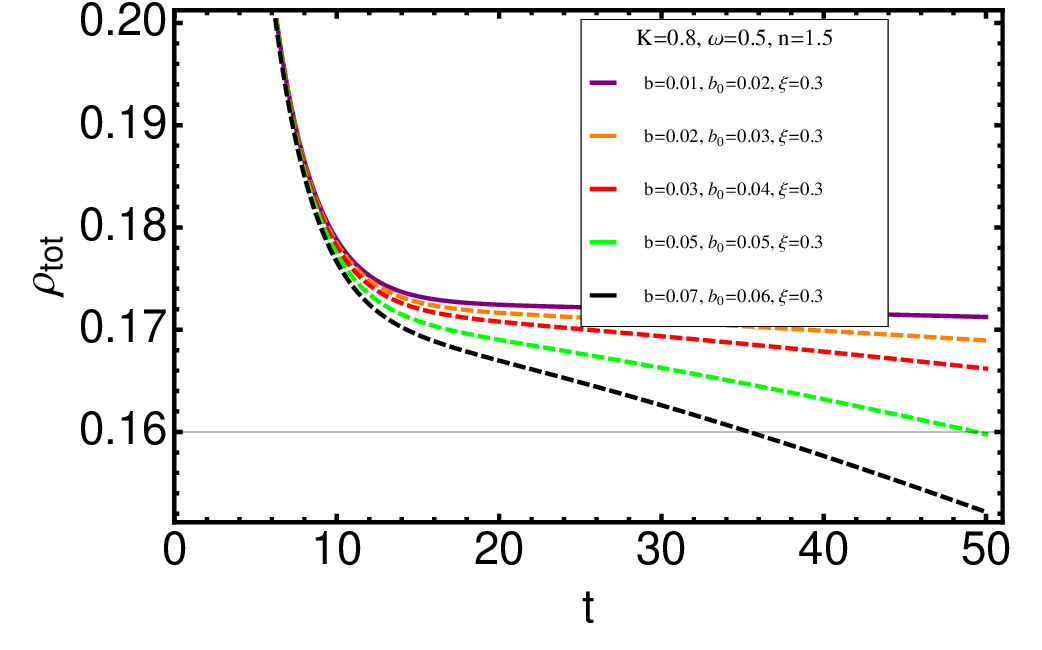} &
\includegraphics[width=50 mm]{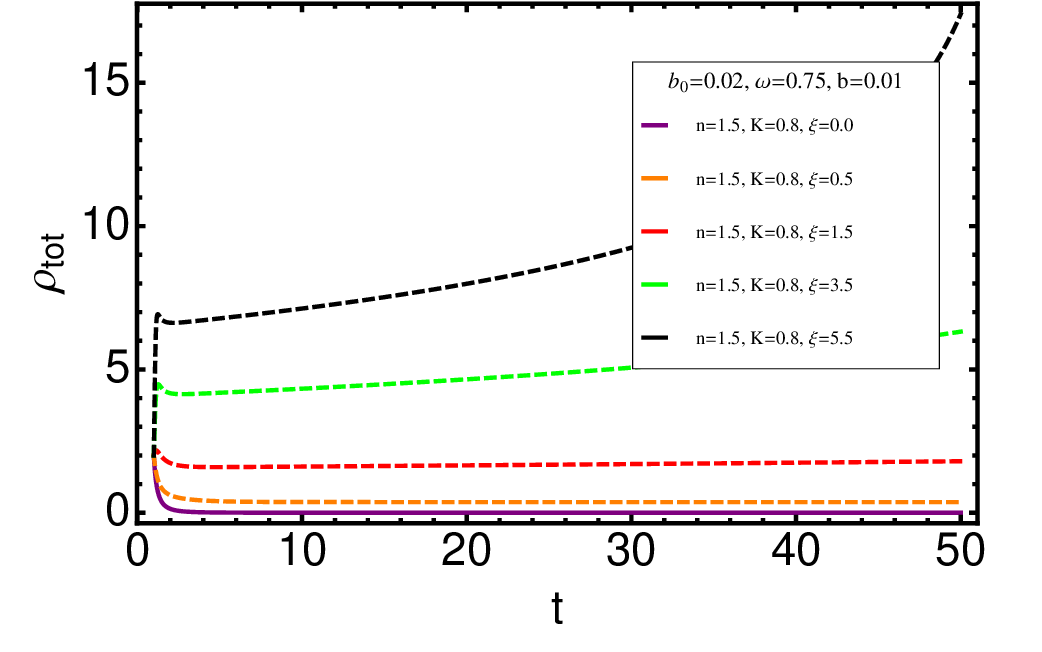}
 \end{array}$
 \end{center}
\caption{Behavior of $\rho_{tot}$ against $t$ for interacting barotropic fluid and viscous palotropic gas.}
 \label{fig:11}
\end{figure}

\begin{figure}[h!]
 \begin{center}$
 \begin{array}{cccc}
\includegraphics[width=50 mm]{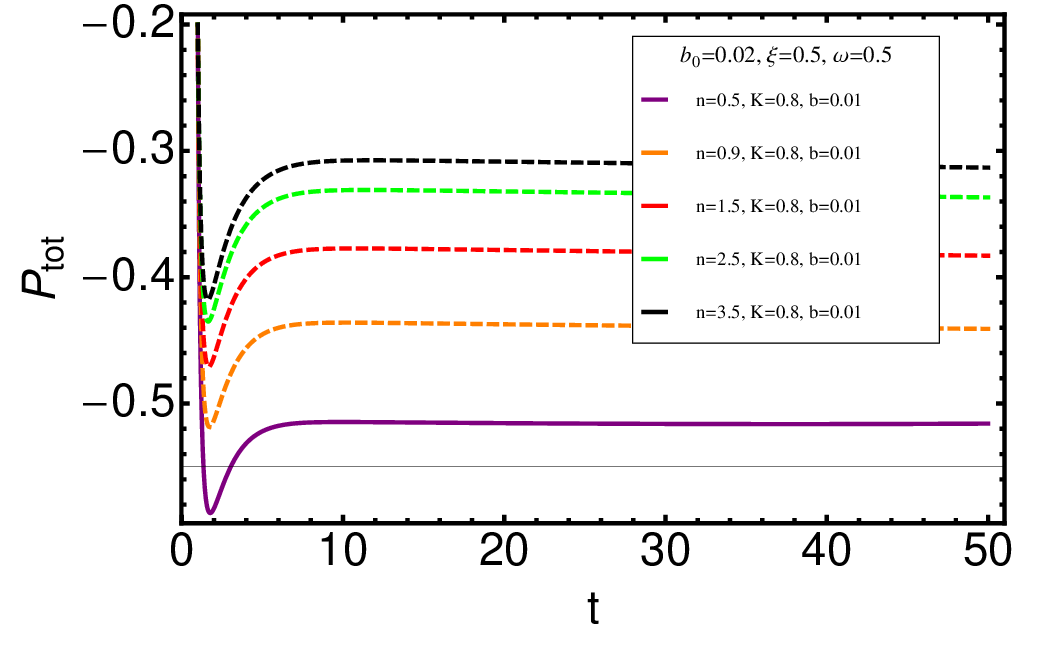} &
\includegraphics[width=50 mm]{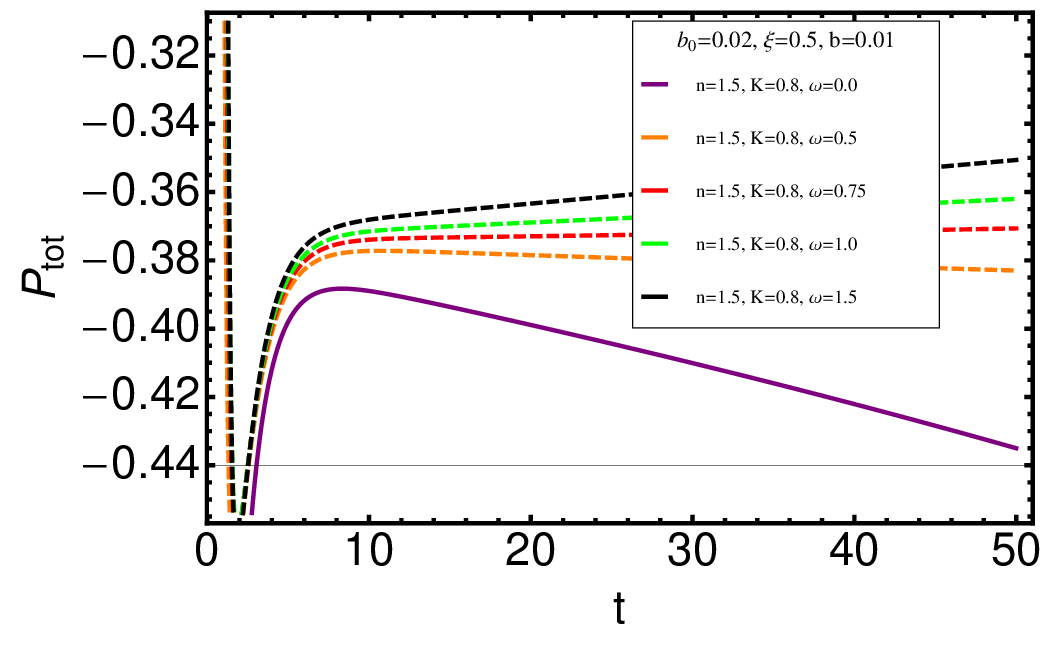}\\
\includegraphics[width=50 mm]{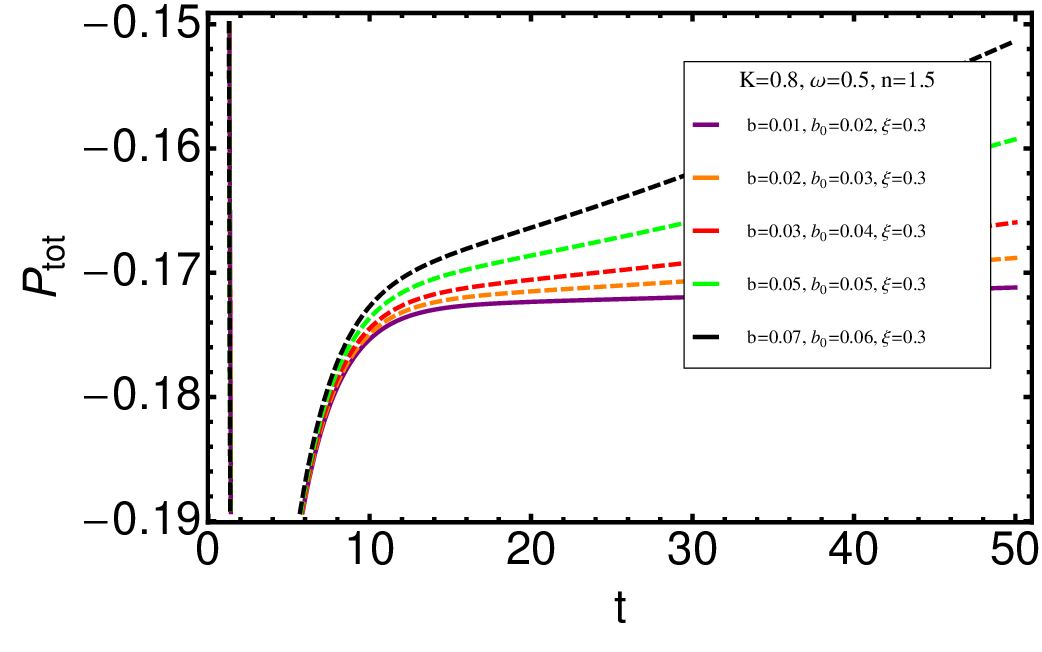} &
\includegraphics[width=50 mm]{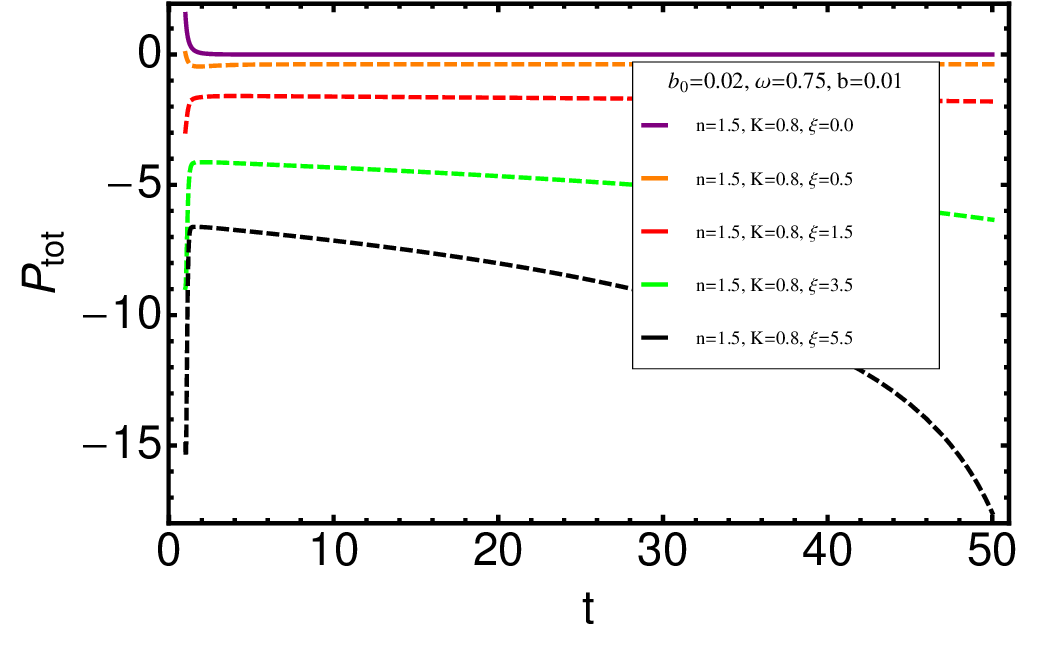}
 \end{array}$
 \end{center}
\caption{Behavior of $P_{tot}$ against $t$ for interacting barotropic fluid and viscous palotropic gas.}
 \label{fig:12}
\end{figure}

\end{document}